\documentclass[course]{houches}
\usepackage[fleqn]{amsmath}
\usepackage{epsfig}
\input{FEYNMAN}

\newcommand{\beq}{\begin{equation}}
\newcommand{\eeq}{\end{equation}}

\newcommand{\Dslash}{/\!\!\!\!D}

\newcommand{\vslash}{/\!\!\!v}

\newcommand{\pslash}{/\!\!\!p}
\newcommand{\kslash}{/\!\!\!k}
\newcommand{\qslash}{/\!\!\!q}

\newcommand{\varepsilonslash}{/\!\!\!\varepsilon}

\newcommand{\bfnabla}{\mbox{\boldmath $\nabla$}}
\newcommand{\bfsigma}{\mbox{\boldmath $\sigma$}}
\newcommand{\bfmu}{\mbox{\boldmath $\mu$}}
\newcommand{{\bfA}}{\bf A}
\newcommand{{\bfk}}{\bf k}
\newcommand{{\bfp}}{\bf p}
\newcommand{{\bfx}}{\bf x}
\newcommand{{\bfy}}{\bf y}
\newcommand{\lsim}{\raisebox{-0.6ex}{
                        $\stackrel{\textstyle <}{\textstyle\sim}$}}

\begin{document}
\setcounter{chapter}{0}
\setcounter{page}{1}
\author[M.B. Wise]{Mark B. Wise}
\address{California Institute of Technology\\ Pasadena, CA  91125
USA\thanks{CALT-68-2172. Work supported in part by the U.S. Dept. of Energy under Grant No.
DE-FG03-92-ER40701}}
\editor{}
\title{}
\chapter{Heavy Quark Physics}

\section{HQET: Physical Motivation}

Heavy quark effective theory (HQET) is the limit of QCD where the
 heavy quark mass, $m_Q$, goes to infinity
with its four-velocity $v^\mu$ held fixed.
This limit is appropriate for physical
systems where there is a heavy quark interacting with light degrees of freedom
that typically carry four-momentum much less than the heavy quark mass.
Consider, for example, a meson with $Q\bar q$ flavor quantum numbers. (Here
$Q$ denotes a heavy quark and $q$ a light quark.)
The size of such a  meson is set by the nonperturbative scale of the
 strong interactions, $r \sim  1/\Lambda_{QCD}$.  Hence, by the uncertainty
principle, the typical momentum carried
by the light degrees of freedom is, $p_\ell \sim \Lambda_{QCD}$.
If an amount of momentum $\delta p_\ell$ is transferred to the heavy
quark the
change in its four-velocity $(p_Q = m_Qv)$ is $\delta v \sim \delta p_\ell/m_Q
\rightarrow 0$, in the infinite mass limit. That is why, for the study
of such hadrons, an appropriate
limit of QCD involves taking $m_Q \rightarrow \infty$ and
holding the heavy quark's four-velocity fixed. As $m_Q \rightarrow \infty$
the strong interactions of the heavy quark become independent of its mass and
spin. Consequently there are new spin-flavor symmetries that arise in this
limit. These approximate symmetries endow HQET with predictive power in the
nonperturbative regime.

\section{HQET: The Effective Field Theory}

The part of QCD Lagrangian containing a heavy quark is
\begin{equation}
{\cal L}_{QCD} = \bar Q (i\Dslash - m_Q)Q.
\end{equation}
where $Q$ is the heavy quark field and $D$ denotes a covariant derivative.
We want to take the limit $m_Q \rightarrow \infty$  with fixed
four-velocity $v^\mu$.  To do this one
scales out the rapidy varying part of the heavy quark field's space-time
dependence by writing,
\begin{equation}
Q = e^{-im_{Q}v\cdot x} Q_v; \quad \vslash Q_v = Q_v.
\end{equation}
Putting this into the QCD Lagrangian
\begin{align}
{\cal L}_{QCD} &\rightarrow \bar Q_v (m_Q \vslash + i \Dslash - m_Q) Q_v
= \bar Q_v i\Dslash Q_v\notag\\
&= \bar Q_v \left (\frac{\vslash + 1}{2}\right) i\Dslash \left(\frac{\vslash +
1}{2}\right) Q_v\notag \\
& = \bar Q_v \left[iv \cdot D + i\vslash \left(\frac{-\vslash + 1}{2}\right)
\left(\frac{\vslash + 1}{2} \right)\right] Q_v.
\end{align}
So in this limit the QCD Lagrange density becomes~\cite{eichten}
\begin{equation}
{\cal L}_{QCD} \rightarrow {\cal L}_{v} = \bar Q_v iv \cdot D Q_v.
\end{equation}
The HQET Feynman rules follow from the above Lagrangian. The heavy quark
propagator is
\begin{equation}
 \frac{i}{v\cdot k + i\varepsilon} \left(\frac{\vslash + 1}{2}\right),
\end{equation}
and the heavy quark gluon interaction vertex is
\begin{equation}
 - igv^\mu T^A,
\end{equation}
where $T^A$ is a color generator.
This can also be derived  by taking the appropriate
limit of the  QCD Feynman rules.
Writing the heavy quark momentum as, $p_Q=m_Qv+k$ , the limit is taken by
neglecting the residual momentum $k$ in comparison with $m_Qv$. For the
propagator this gives,

\begin{equation}
 \frac{i(\pslash_Q + m_Q)}{p_Q^2 - m_Q^2 + i\varepsilon} \rightarrow
 \frac{im_Q (\vslash+1)}{2m_Q v \cdot k + i\varepsilon} = \left(\frac{\vslash +
1}{2}\right)
\left[\frac{i}{v \cdot k + i\varepsilon}\right].
\end{equation}
For the the heavy quark gluon vertex,
\begin{align}
& - igT^A \gamma_\mu \rightarrow - igT^A \left(\frac{\vslash + 1}{2}\right)
\gamma_\mu \left(\frac{\vslash + 1}{2}\right)\notag \\
&= - ig T^A \left[ v^\mu + \gamma_\mu \left(\frac{-\vslash +
1}{2}\right)\right]
\left(\frac{\vslash + 1}{2}\right) = - igT^A v_\mu.
\end{align}
Here we used the fact that the vertex always appears between propagators so
factors of $(\vslash +1)/2$ can be inserted.

 HQET is a theory of the heavy quark. The field $Q_v$ destroys a heavy quark
 but it does not create the corresponding anti-quark. There is no pair
 creation in HQET. This follows straightforwardly from the Feynman rules.
Notice that the propagator only has one pole in $k^0$. So a closed loop
of heavy quark propagators will have all the poles in the zero component of
the loop momentum integration below the real axis. Thus the
integration vanishes.  (When the contour of this integration is
 completed in the upper half plane it encloses no poles.)

We have motivated physically the fixed heavy quark velocity
 superselection rule. Lets now  try to motivate it mathematically.
To derive the effective field theory for a heavy quark of four-velocity
 $v_1$ and a heavy quark of four-velocity $v_2$ write,
\begin{equation}
Q = e^{-m_{q} v_{1} \cdot x} Q_{v_{1}} + e^{-i m_{Q} v_{2} \cdot x}
Q_{v_{2}},
\end{equation}
where $\vslash_1 Q_{v_1}=Q_{v_1}$ and $\vslash_2
Q_{v_2}=Q_{v_2}$. Putting
this in the QCD Lagrangian gives,
\begin{equation}
{\cal L}_{QCD} \rightarrow {\cal L}_{v_{1}}
 + {\cal L}_{v_{2}} +{\cal L}_{v_{1} v_{2}}.
\end{equation}
The cross terms that mix the heavy quark fields of
different velocity is
\begin{align}
{\cal L}_{v_{1} v_{2}}&= e^{- im_{Q} (v_{1} - v_{2}) x} \bar Q_{v_{1}} (m_Q
\vslash_2 + i\Dslash - m_Q) Q_{v_{2}} +h.c.\notag \\
&= e^{-im_{Q} (v_{1} - v_{2})x} \bar Q_{v_{1}} i\Dslash Q_{v_{2}} + h.c.
\end{align}
This contributes a negligible amount to the action in the large mass limit
because of the
rapid variation of the exponential. Similarly the measure in the path integral
factorizes
\begin{equation}
[dQ] \rightarrow [dQ_{v_{1}}] [ dQ_{v_{2}}],
\end{equation}
since the two types of modes are orthogonal.  So we see that
\begin{equation}
{\cal L}_{HQET} = \sum_v \bar Q_v iv \cdot D Q_v.
\end{equation}

For a given four-velocity $v$ the HQET Lagrangian is 
independent of the heavy quark's mass and spin. However, it does
depend on the heavy quark's velocity. All heavy quarks with the same 
four-velocity interact in the same way. If there are $ N$ heavy quarks at
velocity $v$ then the theory has a $SU(2N)$ spin-flavor symmetry~\cite{isgur}.
Note
that this symmetry is a little unusual since it relates states at the same
velocity but different momentum.

Suppose we want to describe an anti-quark instead of a quark.  We start at the
same place except write
\begin{equation}
Q = e^{+im_{Q} \bar v \cdot x} Q_{\bar v} \quad \bar\vslash Q_{\bar v} = -
Q_{\bar v},
\end{equation}
and we anticipate  $Q_{\bar v} $ will be the field that {\it creates} an
anti-quark.  Putting this into the QCD Lagrange density
\begin{align}
{\cal L} &= \bar Q_{\bar v} (- m_Q  \bar\vslash - m_Q + i\Dslash) 
Q_{\bar v}\notag \\
&= \bar Q_{\bar v} i\Dslash Q_{\bar v}\notag \\
&= \bar Q_{\bar v} \left(\frac{1 - \bar\vslash}{2}\right) i\Dslash
\left(\frac{1-\bar\vslash}{2}\right) Q_{\bar v}\notag \\
&= - \bar Q_{\bar v} i \bar v \cdot D Q_{\bar v}.
\end{align}
Heavy anti-quark Feynman rules follow from this or from taking the appropriate
limit of the QCD Feynman rules.  Writing the anti-quark four-momentum
as, $\bar p_Q=-m_Q \bar v-k$, the heavy anti-quark propagator is
\begin{equation}
 \frac{i}{\bar v \cdot k + i\varepsilon} \left(\frac{1 -
\bar\vslash}{2}\right),
\end{equation}
while the heavy anti-quark vertex is
\begin{equation}
+ ig (T^A)^T \bar v_\mu.
\end{equation}
Thus the heavy anti-quark Feynman rules 
are identical to those for the heavy quark
except for the replacement $T^A \rightarrow ( - T^A)^T$.
This is expected since the anti-quark is a color $\bar 3$ and the 
color generators acting on this representation are $(- T^A)^T = T^{A^{*}}$.

\section{$1/m_Q$ Corrections to the HQET Lagrangian}

There are $1/m_Q$ corrections to the HQET Lagrangian.  To
find these lets reexamine the derivation of HQET.  For this we wrote,
\begin{equation}
Q = e^{-im_{Q} v \cdot x} Q_v \quad \vslash Q_v = Q_v.
\end{equation}
Factoring out the phase factor is not an approximation it is just a field
redefinition.  The approximation comes in putting $\vslash Q_v = Q_v$.  So to
go beyond leading order in $1/m_Q$ we make the most general decomposition
\[
Q = e^{-im_{Q} v\cdot x} [Q_v + \chi_v]\]
\begin{equation}
\vslash Q_v = Q_v, \quad \vslash \chi_v = - \chi_v.
\end{equation}
Now the decomposition is completely general.  
The field $\chi$ corresponds partly
to the anti-quark degrees of freedom.  Putting this into the QCD Lagrange
density gives
\begin{align}
{\cal L} &= \bar Q (i\Dslash - m_Q) Q\notag \\
&= [\bar Q_v + \bar \chi_v] (m_Q (\vslash - 1) + i\Dslash) [Q_v + \chi_v]\notag
\\
&= \bar Q_v i v \cdot \Dslash Q_v - \bar \chi_v (iv \cdot D + 2m_Q) \chi_v
\notag \\
&+ \bar Q_v i\Dslash_\perp \chi_v + \bar \chi_v i \cdot \Dslash_\perp Q_v.
\end{align}
It is convenient to introduce a notation for the part of a four-vector
perpendicular to the heavy quark's four-velocity $v$.  For any four-vector $V$
its perpendicular part $V_\perp$ is defined by
\begin{equation}
V_\perp^\mu = V^\mu - (v \cdot V)v^\mu,
\end{equation}
and $v \cdot V_\perp = 0$.  Because $\bar Q_v \vslash \chi_v = \bar \chi_v
\vslash Q_v = 0$ we have replaced $\Dslash$ in the last two terms by
$\Dslash_\perp$.

The field $\chi_v$ has a mass $2m_Q$ and consequently we can integrate it out
of the theory.  At tree level this corresponds to solving the equations of
motion to express
it in terms of $Q_v$.  The field equation
\[
(iv \cdot D + 2m_Q) \chi_v = i \Dslash_\perp Q_v\]
implies that
\begin{equation}
\chi_v = \frac{1}{(iv \cdot D + 2 m_Q)} i \Dslash_\perp Q_v.
\end{equation}
Putting this back into the QCD Lagrange density gives
\begin{eqnarray}
{\cal L} &= \bar Q_v \left\{ i v \cdot D + i\Dslash_\perp \frac{1}{(2m_Q + i v
\cdot D)} i \Dslash_\perp\right\} Q_v\notag \\
&= \bar Q_v iv \cdot D Q_v + \frac{1}{2m_Q} \bar Q_v i \Dslash_\perp
i\Dslash_\perp Q_v + O \left(\frac{1}{m^2_Q}\right). 
\end{eqnarray}
The relationship between the heavy quark field in QCD and in the effective
theory is
\begin{equation}
Q  = e^{-m_{Q} v \cdot x} \left[1 + \frac{i\Dslash_\perp}{2m_Q}\right] Q_v +O
\left(\frac{1}{m^2_Q}\right).\qquad \qquad
\end{equation}
You will often see these formulas written without the $\perp$ subscript.  That
is because if one works at order $1/m_Q$ the equation of motion $v
\cdot D Q_v = 0$ allows one to write $D_\perp$ as a full $D$.  Expanding in
powers of $1/m_Q$,
\[
{\cal L} = {\cal L}_0 + {\cal L}_1 + ...,\]
where the leading term is the HQET Lagrange density
\begin{equation}
{\cal L}_0 = {\cal L}_{HQET} = \bar Q_v i v \cdot D Q_v.
\end{equation}
To simplify ${\cal L}_1$ write,
\begin{eqnarray}
\Dslash_\perp \Dslash_\perp &= \gamma^\mu \gamma^\nu D_{\perp\mu} D_{\perp\nu}
= \frac{1}{2} \{\gamma^\mu, \gamma^\nu\} D_{\perp\mu} D_{\perp\nu} +
\frac{1}{2}
[\gamma^\mu, \gamma^\nu] D_\mu D_\nu\notag \\
&= D_\perp^2 + \frac{1}{4} [\gamma^\mu, \gamma^\nu] [D_\mu, D_\nu].
\end{eqnarray}
Since this is sandwiched between $\bar Q_v$ and $Q_v$ the $D_\perp$'s can be
replaced by $D$'s in the anticommutator term.
Using $\sigma^{\mu\nu} = \frac{i}{2} [\gamma^\mu, \gamma^\nu]~ {\rm and}~ ig
G_{\mu\nu} = [D_\mu, D_\nu],$ the above becomes,
\begin{equation}
\Dslash_\perp \Dslash_\perp = D_\perp^2 + \frac{g}{2} \sigma^{\mu\nu}
G_{\mu\nu}.
\end{equation}
Using this in eq. (3.6) gives the order $1/m_Q$ term in the Lagrange
density~\cite{eichten2}.
\begin{equation}
{\cal L}_1 = - \bar Q_v \frac{D_\perp^2}{2m_Q} Q_v - g \bar Q_v
\frac{\sigma_{\mu\nu}G^{\mu\nu}}{4m_Q} Q_v.
\end{equation}
The first term in ${\cal L}_1$ is the heavy quark kinetic energy.  It breaks
the
flavor symmetry but not the spin symmetry.  The second term is a magnetic
moment interaction $\bfmu_Q \cdot {{\bf B}}_c$.  It breaks both the spin and
flavor symmetries.

This is the Lagrange density at tree level.  The coefficients change when one
includes perturbative $\alpha_s$ connections.  Most importantly the operator
$\bar Q_v \sigma_{\mu\nu}Q^{\mu\nu} Q_v$ requires renormalization so its coefficient
develops a factor $a(\mu)$, where the subtraction point dependence cancels that
of the operator.  The tree level matching of QCD onto HQET that we have
performed so far implies that
\begin{equation}
a (m_Q) = 1 + O(\alpha_s (m_Q)).
\end{equation}
In dimensional regularization with minimal subtraction the kinetic 
energy operator does not get renormalized to all orders in
perturbation theory and its coefficient stays equal to unity.  One way to
understand this is as a consequence of
reparametrization invariance.

\section{Reparametrization Invariance}

A heavy quark's four-momentum can be written as the sum of $m_Qv$ and the residual momentum $k$,
\begin{equation}
p_Q = m_Q v + k.
\end{equation}
This decomposition is not unique.  Typically $k \sim O
(\Lambda_{QCD})$ and a small change in $v$ of order $\Lambda_{QCD}/m_Q$ can be
compensated by a change in $k$ of order $\Lambda_{QCD}$.  Explicitly
\begin{align}
v &\rightarrow v + \varepsilon/m_Q,\notag \\
k &\rightarrow k - \varepsilon,
\end{align}
leaves the heavy quark's four-momentum unchanged.  Since the four-velocity
satisfies $v^2 = 1$ we must, at linear order in
$\varepsilon$, choose
\begin{equation}
v \cdot \varepsilon  = 0.
\end{equation}
We also want to preserve the constraint
\begin{equation}
\vslash Q_v = Q_v.
\end{equation}
Therefore if
\begin{equation}
Q_v \rightarrow Q_v + \delta Q_v,
\end{equation}
then $\delta Q_v$ satisfies
\begin{equation}
\left(\vslash + \frac{\varepsilonslash}{m_Q} \right) (Q_v + \delta Q_v) = Q_v +
\delta Q_v,
\end{equation}
which implies that
\begin{equation}
(1 - \vslash) \delta Q_v = \frac{\varepsilonslash}{m_Q} Q_v.
\end{equation}
Hence we can choose  $\delta Q_v = \frac{\varepsilonslash}{2m_Q} Q_v$, which
satisfies  $\vslash \delta Q_v = - \delta Q_v$.  Therefore the freedom to
choose different heavy quark velocities $v$ implies that the theory should is
invariant under~\cite{luke1}
\begin{align}
v & \rightarrow v + \varepsilon/m_Q\notag \\
Q_v &\rightarrow e^{i\varepsilon \cdot x} \left( 1 +
\frac{\varepsilonslash}{2m_Q}\right) Q_v.
\end{align}
Under this transformation the leading order Lagrange density
\begin{equation}
{\cal L}_0 = \bar Q_v iv \cdot D Q_v,
\end{equation}
transforms to
\begin{equation}
{\cal L}_0 \rightarrow \bar Q_v \left(1 + \frac{\varepsilonslash}{2m_Q}\right)
\left( iv \cdot D + \frac{i\varepsilon \cdot D}{m_Q}\right) \left(1 +
\frac{\varepsilonslash}{2m_Q}\right) Q_v,
\end{equation}
implying that
\begin{equation}
\delta {\cal L}_0  = \frac{1}{m_Q} \bar Q_v (i\varepsilon\cdot D_\perp)
 Q_v.
\end{equation}
Here we used $\bar Q_v \varepsilonslash Q_v = 0$ which follows from  the
constraint $\varepsilon \cdot v = 0$.  The change in ${\cal L}_1$ is
\begin{equation}
\delta {\cal L}_1 \rightarrow - \bar Q_v i 2\frac{\varepsilon \cdot
D_\perp}{2m_Q} Q_v =  - \frac{1}{m_Q} \bar Q_v i \varepsilon \cdot D_\perp Q_v.
\end{equation}
Therefore ${\cal L}_0 + {\cal L}_1$ is invariant under the transformantion
in eq. (4.8) provided the coefficient of the 
kinetic energy term is $1$. Hence if
a regulator is used that preserves invariance under reparametrization transformations the kinetic energy cannot get renormalized.

We have been developing formalism for a $\Lambda_{QCD}/m_Q$ expansion of
physical quantities.   The quarks that are heavy enough for this to be useful
are the charm,
bottom and top which have masses
 $m_c \sim 1.4~{\rm GeV}, m_b \sim 4.8~{\rm GeV}$, and $m_t
\sim 175~{\rm GeV}.$  However $m_t$ is so heavy that it decays before forming a
hadron
($t \rightarrow b + W$, with a width of a few GeV).  So the tools I am
developing here are only useful for the charm and bottom quarks.

\section{Spectrum of Hadrons Containing a Single Heavy Quark}

In the $m_Q \rightarrow \infty$ limit hadrons containing a single heavy quark
are classified (in their rest frame) not only by their total spin (i.e., total
angular momenta) ${{\bf S}}$ but also by the total spin of the light degrees of
freedom ${{\bf S_\ell}}$~\cite{isgur2},
\begin{equation}
{{\bf S}} = {{\bf S}}_\ell + {{\bf S}}_Q.
\end{equation}
For any angular momenta ${{\bf J}}$ its square is  ${{\bf J}}^2 = j(j + 1)$.
Since $s_Q = \frac{1}{2}$ hadrons with a given spin of the light degrees of
freedom $s_\ell$ come in degenerate doublets with spins
\begin{equation}
s = s_\ell \pm \frac{1}{2}.
\end{equation}
In the case of mesons with $Q \bar q$ flavor quantum numbers two doublets
have been observed in the $Q = c$ and $Q = b$ cases.  The ground state doublet
has
$s_\ell = \frac{1}{2}$ and negative parity.  For $Q = c$ these spin $0$ and
spin $1$
mesons are the $D$ and $D^*$ while for 
$Q = b$ they are the $B$ and $B^*$ mesons.
The other observed doublet has $s_\ell = \frac{3}{2}$ and positive parity.  For
$Q = c$
the spin $1$ and spin $2$ members of this doublet are called the $D_1$ and
$D_2^*$ respectively.

There is a very useful phenomenological model that describes with surprising
accuracy the hadronic spectrum.  It is called the nonrelativistic constituent
quark model.  In this model the light $u, d$ quarks have masses of 350 MeV and
the strange quark has a mass of 500 MeV.  The quarks are nonrelativistic and
interact via a potential that is linear at large distances.  The quarks in the
nonrelativistic quark model are not the ones destroyed and created  by the
fields in the QCD Lagrangian.  They are quasi particles and their large masses
arise from nonperturbative strong interaction dynamics.  In this model the
ground state $Q \bar q$ meson has the light anti-quark in an $S$-wave and the
total angular momentum of the light degrees of freedom comes from the light
quark spin.  Consequently this doublet has $s_\ell = \frac{1}{2}$ and negative
parity coming from the intrinsic parity of an anti-quark.  The first
excitations correspond to giving the light anti-quark one unit of orbital
angular
momentum $L = 1$.  Combining this orbital angular momentum of the light degrees
with the spin of the anti-quark gives the possibilities,
$s_\ell = \frac{1}{2}$ and $\frac{3}{2}$.  The parity is now positive because
of the unit
of orbital angular momentum.  As was mentioned before only the $\frac{3}{2}$ doublet
has been observed.

Heavy quark symmetry and a little phenomenological ``lore'' helps provide us
with an understanding of why the other, $s_\ell = \frac{1}{2}$, doublet is not
observed.  The excited mesons in the $s_\ell = \frac{1}{2}$ and $\frac{3}{2}$
doublets decay
to the ground state mesons with emission of a single pion.  The available phase
space is
about 450 MeV and decay to more pions is probably suppressed.  Parity
invariance implies that the one pion decay occurs in an even  partial wave,
either  $L = 0$ or $2$.  The decays $D_2^* \rightarrow D \pi$ and $D_2
\rightarrow D^* \pi$, are $L = 2$ while decay $D_1 \rightarrow D^*\pi$ could
occur in either $L = 2$ or $L = 0$.  But in the $m_c \rightarrow \infty$ limit
the $D_1$ and $D_2^*$ total widths must be equal, so the $L = 0$ amplitude is
forbidden.  On the other hand the $s_\ell = \frac{1}{2}$ doublet of excited
charmed
mesons (denoted $D_0^*$ and $D_1^*$) decay to $D^{(*)} \pi$ in an
$S$-wave.  Amplitudes that go through high partial waves are smaller than those
that go through lower partial waves (they are suppressed by
$(p_\pi/1{\rm GeV})^{2L+1}$)  and so the $D_1$ and $D_2^*$ are narrow having
widths around 20 MeV. The members of the excited $s_\ell = \frac{1}{2}$
doublet decay in an $S$-wave and are expected to be too difficult to observe
because they are quite broad (i.e., widths greater than $\sim 100~{\rm MeV}$).

\section{$v \cdot A = 0$ Gauge}

Calculations in HQET can be performed in almost any gauge.  However in $v \cdot
A = 0$ gauge HQET perturbation theory is singular.  Consider tree level $Qq$
elastic scattering in the rest frame $v = v_r = (1, {\bf 0})$.  In HQET an
on-shell heavy quark has a four-velocity $v$ and a residual momentum $k$ that
satisfies $v \cdot k = 0$.  Suppose the initial heavy quark has zero residual
momentum and the final has residual momentum $k = (0, {\bf k})$.  The tree
level Feynman diagram shown below in Fig. (1) gives the 
heavy quark light quark scattering amplitude
\begin{equation}
{\cal M} = g^2 \bar u(v_r) T^A u(v_r)
 \frac{i}{{\bf k}^2} \bar u(k) T^A \vslash u(0)
\end{equation}
in Feynman or Landau gauge (here $\bar u(k) \kslash u(0) = 0$ was used).  However in
$v \cdot A = 0$ gauge the gluon propagator is
\begin{equation}
\frac{-i}{k^2 + i\varepsilon} \left[g_{\mu\nu} - \frac{1}{v \cdot k} (k_\mu
v_\nu + v_\mu
k_\nu) + \frac{1}{(v \cdot k)^2} k_\mu k_\nu\right].
\end{equation}
The heavy quark kinetic energy cannot be treated as a perturbation in this
gauge because this implies that $v \cdot k = 0$ and the square brackets are ill
defined.
Including the heavy quark kinetic energy in the Lagrangian the residual
momentum of the outgoing heavy quark becomes $k^\mu = ({\bf k}^2/2m_Q, {\bf
k})$ and $v \cdot k =
{\bf k}^2/2m_Q$ is not zero.   At leading order it is the QQA vertex
from the heavy quark kinetic energy that contributes to elastic $Qq$ scattering
in $v \cdot A = 0$ gauge~\cite{luke2}.  The Feynman rule for the $Q_v
(k')\rightarrow Q_v (k) + $ gluon vertex arising from an insertion of the
kinetic energy
operator is, $(g/2 m_Q) (k_\perp + k'_\perp)_\mu = (g/2 m_Q) (k + k')_\mu +
(g/2
m_Q) v \cdot (k + k') v_\mu$.   The part proportional
to $v_\mu$ doesn't contribute, and since $\bar u(k) \kslash u(0) = 0$ only the second
term in the numerator of the propagator matters.   For large $m_Q$ it
reproduces the amplitude ${\cal M}$ above.

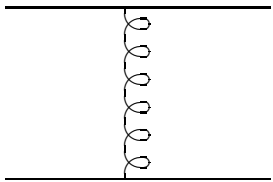
\begin{figure}[h]
\begin{center}
\begin{picture}(10000,10000)
\drawline\fermion[\E\REG](0,0)[10000]
\drawline\gluon[\N\REG](4500,0)[6]
\THICKLINES
\drawline\fermion[\E\REG](0,\gluonbacky)[10000]
\end{picture}
\end{center}
\vskip 0.3in
\caption{Qq scattering via one gluon exchange. }
\end{figure}

For the calculation of on-shell amplitudes in $v \cdot A = 0$ gauge the heavy
quark kinetic energy must be considered as
leading and it is the QQA vertex from this operator that gives rise to
the leading $Qq$ scattering amplitude.  However, for off-shell Green
functions this gauge can be used in HQET, with the kinetic energy treated as a
perturbation.  Since anomalous dimensions follow from such Green functions we
can conclude, for example, that the
anomalous dimension of the operator $\bar Q_v \Gamma Q_v$ vanishes, without
performing any calculation. There is another way to understand this. The 
operator $\bar Q_v \Gamma Q_v$ is a charge density of the heavy quark
spin-flavor symmetry and therefore cannot be renormalized.

\section{Mass Formulae}

The masses of hadrons containing a single heavy quark can be expanded in powers
of $1/m_Q$.  In the rest frame $v = v_r = (1,{\bf 0})$
\begin{align}
M_{H^{(Q)}} = m_Q + \bar\Lambda &+ \frac{\langle H^{(Q)}_{v_r} | \bar Q_{v_{r}} 
D_\perp^2 Q_{v_{r}} | H^{(Q)}_{v_r}\rangle}{2 \cdot 2m_Q}\notag \\
&+ a (\mu) \frac{\langle H^{(Q)}_{v_r} |\bar Q_{v_{r}} g \sigma_{\alpha\beta}
G^{\alpha\beta}Q_{v_{r}}|H^{(Q)}_{v_r}\rangle}{2 \cdot 4m_Q} +\ldots,
\end{align}
where the ellipses denote terms suppressed by more powers of $1/m_Q$.
$\bar\Lambda$ is the energy of the light degrees of freedom in the $m_Q
\rightarrow \infty$ limit.  It is the same for both members of a spin symmetry
doublet but of course is different for different doublets.  The kinetic energy
term is also the same for both members of a given doublet and it is
conventional to write
\begin{equation}
2 \lambda_1 = - \langle H^{(Q)}_{v_r} | \bar Q_{v_{r}} D_\perp^2
Q_{v_{r}}|H^{(Q)}_{v_r}\rangle,
\end{equation}
$\lambda_1$ is independent of the heavy quark mass and is of order
$\Lambda_{QCD}^2$.  The matrix elements in eq. (7.1) are taken in HQET
with the hadron states having zero residual momentum. 
There is a factor of two in the denominator
because of the HQET normalization convention, 
$\langle H^{(Q)}_{v'} (\vec k') | H^{(Q)}_v (\vec k) \rangle = 
2 v^0 (2 \pi)^3 \delta_{vv'} \delta^3 (\vec k - \vec k')$.  
Since $\bar Q_{v_{r}} \sigma_{\alpha\beta}
G^{\alpha\beta} Q_{v_{r}} = \bar Q_{v_{r}} \bfsigma \cdot {\bf B}_c
Q_{v_{r}}$ the matrix element of this operator is proportional to ${\bf S}_Q
\cdot {\bf S}_\ell$.  It is conventional to write
\begin{equation}
({\bf S}_Q \cdot {\bf S}_\ell) 16 \lambda_2 =
 - a (\mu) \langle H^{(Q)}_{v_r} | \bar
Q_{v_{r}} g \sigma_{\alpha\beta} G^{\alpha\beta} Q_{v_{r}}|H^{(Q)}_{v_r}\rangle.
\end{equation}
We can simplify this using
\begin{align}
{\bf S}_Q \cdot {\bf S}_\ell &= ({\bf S}^2 - {\bf S}_Q^2 - {\bf
S}_\ell^2)/2\notag \\
&= (s (s+1) - 3/4 - s_\ell (s_\ell + 1))/2.
\end{align}
For the higher spin $(s_+)$ member of a doublet $s_\ell = s_+ - 1/2$ and the
above is
\begin{align}
2 ({\bf S}_Q \cdot {\bf S}_\ell)_+&= s_+(s_+ + 1) - 3/4 - (s_+ - 1/2) (s_+ +
1/2)\notag \\
&= s_+^2 + s_+ - 3/4 - s_+^2 + 1/4\notag \\
&= \frac{1}{2} (s_+ - 1)  = \frac{1}{2} (2s_- + 1),
\end{align}
while for the lower spin $(s_-)$ member of a doublet
\begin{align}
2 ({\bf S}_Q \cdot {\bf S}_\ell)_- &= s_- (s_- + 1) - 3/4 - (s_- + 1/2) (s_- +
3/2)\notag \\
&= s_-^2 + s_- -3/4 - s_-^2 - 2s_- - 3/4\notag \\
&= - \frac{1}{2} \cdot (2s_- + 3) = - \frac{1}{2} (2s_+ +1).
\end{align}
So
\begin{equation}
({\bf S}_Q \cdot {\bf S}_\ell)_\pm = \pm \frac{1}{4} n_\mp.
\end{equation}
where $n_\mp = (2s_\mp + 1)$ is the number of spin components.  The matrix
element $\lambda_2$ is defined
\begin{equation}
\lambda_2 = \frac{\pm 1}{4 \cdot n_{\mp}} a(\mu) \langle H^{(Q)}_{(\pm)}|
\bar Q_{v_{r}} g \sigma_{\alpha\beta} G^{\alpha\beta} Q_{v_{r}} | 
H^{(Q)}_{(\pm)}
\rangle,
\end{equation}
and the mass formula is
\begin{equation}
m_{H^{(Q)}_{(\pm)}} = m_Q + \bar\Lambda - \frac{\lambda_1}{2m_Q}  \pm
\frac{n_\mp
\lambda_2}{2m_Q} +\ldots.
\end{equation}
Note that the spin averaged mass
\begin{equation}
\bar m_{H^{(Q)}} = \frac{n_+ m_{H^{(Q)}_{(+)}} + n_- m_{H^{(Q)}_{(-)}}}{n_+ + n_-},
\end{equation}
is independent of $\lambda_2$.  $\lambda_1$ is independent of the heavy quark
mass and $\lambda_2$ has a weak logarithmic dependence on $m_Q$. 
 The difference in mass
between the two members of a doublet determines $\lambda_2$.
\begin{equation}
m_{H^{(Q)}_{(+)}} - m_{H^{(Q)}_{(-)}} = \frac{(n_- + n_+)}{2m_Q} \lambda_2
= \frac{(2s_\ell + 1)}{m_Q} \lambda_2.
\end{equation}
For the ground state multiplet of mesons with $s_\ell = \frac{1}{2}$ the $B^* -
B$
splitting implies that $\lambda_2 (m_b) = 0.12~{\rm GeV}^2$ while the $D^* - D$
splitting gives $\lambda_c (m_c) = 0.10~{\rm GeV}^2$.
  Here (and throughout these
lectures) I use $m_b = 4.8~{\rm GeV}$ and $m_c = 1.4~{\rm GeV}$.  The charmed states in
the $s_\ell^{\pi_{\ell}} = \frac{3}{2}^+$ multiplet are the $D_2^* (2460)$ and
$D_1
(2420)$.  The mass splitting between these states implies  that for this
multiplet $\lambda'_2 = 0.013~{\rm GeV}^2$.  In these lectures I use the notation
$\bar\Lambda, \lambda_1, \lambda_2, \bar m_P$ for the ground state meson
multiplet and $\bar\Lambda', \lambda'_1, \lambda'_1, \bar m'_P$ for the excited
$s_\ell^{\pi_{\ell}} = \frac{3}{2}^+$ doublet.

The values of $\bar\Lambda, \lambda_1$ are not well known.  A
determination from the lepton energy spectrum in inclusive semileptonic $B$
decay gives  $\bar\Lambda \simeq 0.4~{\rm  GeV},
 \lambda_1 \simeq - 0.2~ {\rm GeV}^2$, but the uncertainties are very large~\cite{gremm}.

Using the mass formula above
it is straightforward to deduce that~\cite{Leibovich}
\begin{align}
\bar\Lambda' - \bar\Lambda &= \frac{m_b (\bar m_B' - \bar m_B) - m_c (\bar
m_D')}{m_b - m_c} \simeq 0.39~{\rm GeV} \notag \\
\lambda'_1 - \lambda_1 &= \frac{2m_c m_b [(\bar m_B' - \bar m_B) - (\bar m_D' -
\bar m_D)]}{m_b - m_c} \simeq - 0.23~{\rm GeV}^2.
\end{align}
Here I used $\bar m_B' = 5.73~{\rm GeV}$~\cite{feindt}.  There is considerable uncertainty in this value
because the peak of the $B^* \pi$ mass distribution measured at LEP may not
correspond to the narrow $\frac{3}{2}^+$ doublet.  The above formula for
$\bar\Lambda' - \bar\Lambda$ holds up to corrections of order
$\Lambda_{QCD}^3/m_Q^2$ since
the dependence on $\lambda_1$ cancels out between the two terms in the
numerator.  The extraction of
$\bar\Lambda' - \bar\Lambda$ is not very sensitive to the numerical choices for
$m_{b,c}$.

\section{Covariant Representation of Fields}

The fields in the ground state meson doublet are denoted by $P^{(Q)}_v$ and
$P_{\mu v}^{*(Q)}$.  The vector meson field satisfies 
$v^\mu P_{\mu v}^{*(Q)} = 0$.  It
is convenient to combine the $P^{(Q)}_v$ and $P_{\mu v}^{*(Q)}$ fields into a single
object
that transforms in a simple way under heavy quark spin transformations.
The mesons are composed of a heavy quark $Q$ and light degrees of
freedom as is the bispinor $Q_\alpha \bar\ell_\beta$.  So the object we want
should transform in this way under Lorentz transformations.  The obvious choice
is~\cite{bj,falk},
\begin{equation}
H_v^{(Q)} =\frac{(1 + \vslash)}{2} [\gamma_\mu P_v^{*\mu(Q)} - \gamma_5
P_v^{(Q)}].
\end{equation}
The bispinor field satisfies
\begin{equation}
\vslash H_v^{(Q)} = H_v^{(Q)} ~{\rm and}~ H_v^{(Q)} \vslash = - H_v^{(Q)},
\end{equation}
and under heavy quark spin transformations
\begin{equation}
H_v^{(Q)} \rightarrow D(R)_Q H_v^{(Q)}.
\end{equation}
The matrix $D(R)_Q$ is the usual Dirac four-component spinor representation of
rotations and it satisfies $\gamma^0 D^\dagger (R)_Q \gamma^0 =
D(R)_Q^{-1}$.  The $\gamma_5$ in eq. (8.1) is dictated by parity, under
which $H_v^{(Q)}$ transforms as
\[
H_v^{(Q)} (x) \rightarrow \gamma^0 H_{v_{p}}^{(Q)} (x_P) \gamma^0,\]
where
\begin{equation}
x_P = (x^0, - {\bf x}), \qquad v_P  = (v^0, - {\bf v}).
\end{equation}
Note that
\begin{equation}
\bar H_v^{(Q)} = [\gamma^\mu P_{v\mu}^{*(Q)\dagger} + \gamma_5
P_v^{(Q)^\dagger}] \frac{(1 + \vslash)}{2}
\end{equation}
and that
\begin{equation}
Tr \bar H_v^{(Q)} H_v^{(Q)} = - 2P_v^{(Q)\dagger} P_v^{(Q)} +
2P^{*(Q)\dagger}_{v\mu} P_v^{*(Q)\mu},
\end{equation}
is invariant under Lorentz transformations and heavy quark spin symmetry.

\section{Semileptonic $B \rightarrow D^{(*)} e\bar\nu_e$ Decay in the $m_Q
\rightarrow \infty$ Limit}

In semileptonic $K \rightarrow \pi e\bar\nu_e$ decay it is conventional to use
the variable $q^2 = (p_e + p_{\nu_{e}})^2$ to describe the kinematics of the
decay.  For $B \rightarrow D^{(*)} e\bar\nu_e$ decay $q^2 = (p_e +
p_{\nu_{e}})^2 = (p_B - p_{D^{(*)}})^2$ is in the range  $0 \leq q^2  \leq (m_B
- m_{D^{(*)}})^2$.  The
maximum value of $q^2$ corresponding to the zero recoil kinematic point where
the $D^{(*)}$ is at rest in the rest frame of the $B$ meson.  The minimum value
of $q^2$ occurs when the $D^*$ is recoiling maximally.  Even though $q^2$ can
be of order $m_{c,b}^2$ the momentum transfer to the light degrees of freedom
is much less than the heavy quark masses.  The momentum transfer to the light
degrees of freedom is of order
\begin{equation}
q^2_{light} \sim (\Lambda_{QCD} v' - \Lambda_{QCD}v)^2 = 2 \Lambda_{QCD}^2
(v \cdot v' - 1).
\end{equation}
For $B \rightarrow D^{(*)} e\bar\nu_e$ decay a better variable than $q^2$ to
use is the dot product of $B$ and $D^{(*)}$ four-velocities
$w = v \cdot v'$.  It is related to $q^2$ by,
\begin{equation}
w = v \cdot v' = \frac{m_B^2 + m_{D^{(*)}}^2 - q^2}{2m_B m_{D^{(*)}}}.
\end{equation}
For these decays $1< w \lsim 1.6$ and hence
 $q_{light}^2$ is small compared to the
heavy quark masses over the entire phase space.  Consequently HQET is the
appropriate starting point.  The invariant matrix element for the semileptonic
decay is
(since the electron and neutrino don't interact strongly)
\begin{equation}
{\cal M} = \frac{G_F}{\sqrt{2}}V_{cb} \langle D^{(*)} (p') |\bar c \gamma_\mu (1 - \gamma_5)
b| B(p)\rangle [\bar u (p_e) \gamma^\mu (1 - \gamma_5) v (p_{\nu_{e}})].
\end{equation}
All the complicated nonperturbative strong interaction physics is in matrix
elements of the vector current $V_\mu = \bar c \gamma_\mu b$ and axial vector
current
$A_\mu = \bar c \gamma_\mu \gamma_5 b$.  These matrix elements can be written
in terms of Lorentz scalar form factors and it is convenient for comparisons
with the predictions of HQET to write these form factors as functions of $w = v
\cdot v'$, where the $B$ meson has four-momentum $p = m_B v$ and the $D^{(*)}$
meson has four-momentum $p' = m_{D^{(*)}} v'$.  When we match onto
HQET we use this choice of four-velocities for the
heavy quark fields and the heavy meson states. This 
corresponds to setting the residual
momentum of the $B$ and $D^{(*)}$ states to zero. 
In terms of Lorentz scalar form factors the matrix elements are
\begin{align}
\frac{\langle D(p')|V^\mu |B(p)\rangle}{\sqrt{m_B m_D}} &= h_+ (w) (v + v')^\mu
+ h _- (w) (v - v')^\mu\notag \\
\frac{\langle D^* (p',\varepsilon) | V^\mu| B(p)\rangle}{\sqrt{m_B m_{D^{*}}}}
&=ih_V (w) \varepsilon^{\mu\nu\alpha\beta} \varepsilon_\nu^* v_\alpha'
v_\beta\notag \\
\frac{\langle D^* (p', \varepsilon) | A^\mu | B(p)\rangle}{\sqrt{m_B
m_{D^{*}}}}
& = h_{A_{1}} (w) (w + 1) \varepsilon^{*\mu} - h_{A_{2}} (\varepsilon^* \cdot
v) v^\mu \notag \\
&- h_{A_{3}} (w) (\varepsilon^* \cdot v) v^\mu.
\end{align}

There are several features of these formulas that need explanation.  Firstly
note that for the $B \rightarrow D$ case there is no axial current
matrix element.  This is a consequence of parity invariance of the strong
interactions.  Suppose we write
\begin{equation}
\frac{\langle D(p') | A^\mu |B (p)\rangle}{\sqrt{m_B m_D}} = h_{A_{+}} (w) (v +
v')^\mu + h_{A_{-}} (w) (v - v')^\mu,
\end{equation}
under parity
\begin{align}
P^{-1} \vec A P &= \vec A,\notag \\
P^{-1} A^0 P &= - A^0,
\end{align}
and
\begin{equation}
P |D(p) \rangle = - |D(p_P)\rangle
\end{equation}
where
\begin{equation}
p_P = (p^0, - \vec p).
\end{equation}
Now
\begin{align}
&\langle D(p'_P) | A^\mu| B(p_P) \rangle = h_{A_{+}} (w) (v_P + v'_P)^\mu +
h_{A_{-}} (w) (v_P - v'_P)^\mu \notag \\
& = \langle P D (p') | A^\mu | PB(p)\rangle = \langle D(p') | P^{-1} A^\mu
P|B(p)\rangle.
\end{align}
This implies that
\begin{align}
&- [h_{A_{+}} (w) (v + v')^0 + h_{A_{-}} (w) (v - v')^0] \notag \\
&= h_{A_{+}} (w) (v + v')^0 + h_{A_{-}} (w) (v - v')^0,
\end{align}
and
\begin{align}
&[h_{A_{+}} (w) (v + v')^j + h_{A_{-}} (w) (v - v')^j]\notag \\
& = - h_{A_{+}} (w) (v + v')^j - h_{A_{-}}(w) (v - v')^j.
\end{align}
Consequently
\begin{equation}
h_{A_{+}} = h_{A_{-}} = 0.
\end{equation}
Similarly the $i$ in the matrix element of the vector current between $B$ and
$D^*$ is from time reversal invariance of the strong interactions.  Finally
the factor of $\sqrt{m_B m_{D^{(*)}}}$ in the denominator occurs because of our
normalization of states.  In full QCD we use the standard normalization
\begin{equation}
\langle H(p') | H(p)\rangle = 2 E (2\pi)^3 \delta^3 (\vec p - \vec
p'),
\end{equation}
which has a factor of $m_H$ from $E$.  With $\sqrt{m_B
m_{D^{(*)}}}$ in the denominator the matrix elements on the left hand side become independent of $m_Q$ in the large mass limit.

The first step in using heavy quark symmetry to constrain the form of the
matrix elements is to match the weak current onto an operator in HQET.  Neglecting
$\alpha_s (m_{c,b})$ and $\Lambda_{QCD}/m_{c,b}$ corrections this matching is
\begin{align}
\bar c \gamma_\mu b & = \bar c_{v'} \gamma_\mu b_v\notag \\
\bar c \gamma_\mu \gamma_5 b &= \bar c_{v'} \gamma_\mu \gamma_5 b_v.
\end{align}
Operators $\bar c_{v'} \Gamma b_v$ transform in a particular way under heavy
quark spin symmetry.  If we pretend
that $\Gamma \rightarrow D_c (R) \Gamma D^{-1} (R)_b$ then the operator is
invariant.  For $B \rightarrow D^{(*)}$ matrix elements we want to represent
the operator $\bar c_{v'} \Gamma b_v$ in terms of $B$ and
$D^{(*)}$ fields constructed  so that it transforms in the same way as the
quark operator.  This gives
\begin{equation}
\bar c_{v'} \Gamma b_v = Tr \bar H_{v'}^{(c)} \Gamma H_v^{(b)} X,
\end{equation}
where $X_{\alpha\beta}$ is a general matrix written in terms of $v, v'$ the
gamma matrices and the identity matrix,
\begin{equation}
X = X_0 + X_1 \vslash + X_2 \vslash' + X_3 \vslash \vslash'.
\end{equation}
But since $\vslash H_v^{(b)} = H_v^{(b)}$ and $\bar H_{v'}^{(c)} \vslash' =
-\bar H_{v'}^{(c)}$ each term in eq. (9.16) has the same effect.  We can
equivalently write,
\begin{equation}
X = - \xi (w), 
\end{equation}
where $\xi(w)$ is usually called the Isgur-Wise function.
Taking the $B\rightarrow D^{(*)}$ matrix element of eq. (9.15) implies
the following relations between the form factors~\cite{isgur}
\begin{align}
h_+ (w) &= h_V (w) = h_{A_{1}} (w) = h_{A_{3}} (w) = \xi (w)\notag \\
h_- (w) &= h_{A_{2}} (w) = 0.
\end{align}
At zero recoil $\bar c_v \Gamma b_b$ is a generator of spin flavor
symmetry and its matrix element is known.  This implies that~\cite{isgur,nussinov}
\begin{equation}
\xi (1) = 1.
\end{equation}

The HQET Feynman rules are close to the full QCD Feynman rules for residual
momentum small compared with $m_Q$. However, in loop integrations the residual
momentum gets arbitrarily large and for regions of integration where $k>m_Q$
the HQET results are very different from those in full QCD. Fortunately QCD is
asymptotically free and therefore these differences can be taken into account
by adding perturbative corrections to the coefficients of HQET operators.
The perturbative corrections to the matching of the weak currents onto HQET
operators do not cause a loss of predictive power since dimensional analysis
dictates that the HQET operators that occur in the matching condition are
of the form $\bar c_{v'}\Gamma b_v$. The $B \rightarrow D^{(*)}$ matrix elements
of any operator of this type (not just those occuring in eq. (9.14) where
$\Gamma$ is $\gamma_{\mu}$ or $\gamma_{\mu} \gamma_5$ ) are determined by $\xi$.
I will not discuss such perturbative corrections in these lectures. 
An excellent review of this subject, with references to the 
original literature, can by found in ~\cite{neubert}.

\section{Luke's Theorem for the $1/m_Q$ Corrections}

There are two sources of $1/m_Q$ corrections to the form factors, corrections to
the currents and corrections to the states. Neither of them change the matrix elements of the weak currents at zero recoil. This 
result is important for the determination of the $b \rightarrow c$ element of
the Cabibbo-Kobayashi-Maskawa matrix from exclusive $B$ decays and is called
Lukes theorem~\cite{luke}.
 The corrections to the currents arise from the $1/m_Q$ term in
\begin{equation}
Q = e^{-im_{Q}v\cdot x} \left[ 1 + \frac{i\Dslash_\perp}{2m_Q}\right] Q,
\end{equation}
and those from corrections to the states arise from ${\cal L}_1$ in eq. (3.11).  I will discuss both types.  For the currents including the
order $1/m_Q$ term in eq. (10.1) gives
\begin{equation}
\bar c \Gamma b = \bar c_{v'} \Gamma b_v + \bar c_{v'}
\left(-\frac{i\overleftarrow{\Dslash}_\perp}{2m_c} \Gamma + \Gamma \frac{i
{\Dslash}_\perp}{2m_b}\right) b_v,
\end{equation}
for the relationship between the currents in full QCD and HQET. For the
$B \rightarrow D^{(*)}$ matrix elements of the $1/m_Q$ terms
\begin{align}
\bar c_{v'} i \overleftarrow{D}_\mu \Gamma b_v &= Tr \bar
H_{v'}^{(c)} \Gamma H_v^{(b)} M_\mu^{(c)}\notag \\
\bar c_{v'} \Gamma i {D}_\mu b_v &= Tr \bar H_{v'}^{(c)} \Gamma
H_v^{(b)}
M_\mu^{(b)},
\end{align}
where the general form for the matrices $M_\mu^{(Q)}$ is
\begin{align}
M_\mu^{(c)} &= \xi_+^{(c)} (v + v')_\mu + \xi_-^{(c)} (v - v')_\mu -
\xi_3^{(c)} \gamma_\mu\notag \\
M_\mu^{(b)} &= \xi_+^{(b)} (v + v')_\mu + \xi_-^{(b)} (v - v')_\mu -
\xi_3^{(b)} \gamma_\mu.
\end{align}
Here $\xi_+^{(Q)}, \xi_-^{(Q)}$ and $\xi_3^{(Q)}$ are functions of $w$.  But
we have the following identity valid for a $B\rightarrow D^{(*)}$ matrix
element
\begin{equation}
i D_\mu (\bar c_{v'} \Gamma b_v) = \bar\Lambda (v' - v)_\mu \xi (w) Tr \bar
H_{v'}^{(c)} \Gamma H_v^{(b)},
\end{equation}
which implies the relations
\begin{align}
& \xi_-^{(c)} + \xi_-^{(b)} = - \bar\Lambda \xi, \notag \\
& \xi_+^{(b)} + \xi_+^{(c)} = 0, \notag \\
& \xi_3^{(b)} + \xi_3^{(c)} = 0.
\end{align}
Eqs. (10.6) allow us to express $\xi_j^{(b)}$ in terms of $\xi_j^{(c)}$.  The
leading order equations of motion imply that
\begin{equation}
v'_\mu \bar C_{v'} i \overleftarrow{D}^\mu \Gamma b_v = 0, ~{\rm{and}}~ v_\mu
\bar c_{v'} \Gamma iD_\mu b_v = 0.
\end{equation}
This gives the following relations between the form factors
\begin{equation}
\xi_+^{(c)} (1 + w) + \xi_-^{(c)} (1 - w) + \xi_3^{(c)} = 0
\end{equation}
\begin{equation}
\xi_+^{(b)} (1 + w) + \xi_-^{(b)} (1 - w) + \xi_3^{(b)} = 0 .
\end{equation}
So all the current corrections can be expressed in terms of $\bar\Lambda,
\xi (w)$ and $\xi_3 (w)$.   At zero recoil $\xi_-$ has no effect on the matrix
element.  At this kinematic point the above relations become
\begin{align}
2 \xi_+^{(c)} (1) + \xi_3^{(c)} (1) &= 0, \notag \\
2 \xi_+^{(b)} (1) + \xi_3^{(b)} (1) &= 0.
 \end{align}
Eqs. (10.10) imply that at zero recoil (where $v'=v$) $M_{\mu}^{(Q)}$ is
proportional to $\gamma_{\mu} + v_{\mu}$ which vanishes in the trace of
eqs. (10.3). Hence we conclude that
the order $1/m_Q$ corrections to the current do not contribute to the
$B \rightarrow D^{(*)}$ matrix elements of the weak currents at zero recoil.

Now we consider the Lagrangian corrections.  These are corrections to the
relation between heavy meson  states in full QCD to those in HQET.  They are
represented  by the HQET matrix element of
\begin{equation}
i \int d^4 x T \{{\cal L}_1 (x) (c_{v'} \Gamma b_v)
(0)\}.
\end{equation}
There are two types of terms in ${\cal L}_1$.  One is the heavy quark
kinetic energy.  It does not break the spin symmetry and so its effects can
be absorbed into a redefinition of the Isgur--Wise function, $\xi (w)$.
Next consider the charm quark chromomagnetic term. For $B \rightarrow D^{(*)}$
matrix elements,
\[
i\int d^4 x a (\mu) T\{(\bar c_{v'} g\sigma_{\alpha\beta} G^{\alpha\beta} c_{v'})
(x) (\bar c_{v'} \Gamma b_v)(0)\}\]
\[
= Tr \bar H_{v'}^{(c)} \sigma_{\alpha\beta} \left(\frac{1 + \vslash'}{2}\right)
\Gamma H_v^{(b)} X_{(v, v')}^{\alpha\beta},\]
where
\begin{equation}
X^{\alpha\beta} = i X_2 (w) (v^\alpha \gamma^\beta - v^\beta \gamma^\alpha) +
X_3 (w)\sigma^{\alpha\beta}.
\end{equation}
Since
\[
\left(i \int d^4 x a (\mu) T\{(\bar b_v g \sigma_{\alpha\beta} G^{\alpha\beta} b_v) (x)(\bar
b_v \bar\Gamma c_{v'})(0)\}\right)^\dagger\]
\begin{equation}
= - i \int d^4 x a (\mu) T\{(\bar b_v g \sigma_{\alpha\beta} G^{\alpha\beta} b_v
)(x)(\bar c_{v'} \Gamma b_v)(0)\},
\end{equation}
the bottom quark chromomagnetic correction is also characterized by $X_2$ and
$X_3$.  Only $X_3$ can contribute at zero recoil.

These Lagrangian corrections 
also preserve the normalization of matrix elements at
zero recoil.   The $1/m_Q$ corrections change a heavy meson state to
$|P^{(Q)(*)}\rangle + (\varepsilon/m_Q) |S^{(Q)(*)}\rangle$ where the new state
$|S^{(Q)(*)}\rangle$ is orthogonal to the $m_Q \rightarrow \infty$ state
$|P^{(Q)(*)}\rangle$, and $\varepsilon$ is a quantity of order
$\Lambda_{QCD}$.  At zero recoil $\bar c_v \Gamma b_v$ is a charge
of the heavy quark effective theory.  It takes $|P^{(b)(*)}\rangle$ into
$|P^{(c)(*)}\rangle$ and consequently $\langle S^{(c)(*)}| \bar c_v
\Gamma b_v | P^{(b)(*)}\rangle = 0$.  So there are no $\Lambda_{QCD}/m_Q$
corrections at zero recoil.  In other words even 
after absorbing kinetic energy
correction into $\xi$  we still have
normalization condition $\xi(1) = 1$ and furthermore $X_3 (1) = 0$.

\section{Kinematics of Inclusive $B$ Decay}

Semileptonic $B$-meson decays to final states involving a charm quark arise
from matrix elements of the weak Hamiltonian density
\begin{equation}
H_W = \frac{G_F}{\sqrt{2}} V_{cb} \bar c\gamma^\mu (1 - \gamma_5) b
\bar e \gamma_\mu (1 - \gamma_5)\nu_e.
\end{equation}
In 3-body exclusive semileptonic decay one looks at fixed
final states, like $D e\bar\nu_e$.  The differential decay distribution has two
independent kinematic variables which can be taken to be $E_e$ and
$E_{\nu_{e}}$.  The
decay distribution depends implicitly on the mass of the final hadronic state
which is treated as a constant.  In inclusive decays one ignores all
details about the final hadronic state $X_c$ and sums over final states
containing a charm quark.  In addition to the usual kinematic variable $E_e$
and $E_{\nu_{e}}$ that occur in exclusive semileptonic decay there is a third
variable which will be chosen to be $q^2$, the invariant mass of the virtual
$W$-boson.

The differential decay distribution for inclusive semileptonic decay is
\[
\frac{d\Gamma}{dq^2 dE_e dE_{\nu_{e}}} = \int \frac{d^4 p_e}{(2\pi)^3} \int
\frac{d^4p_{\nu_{e}}}{(2\pi)^3} \delta (p^2_{\nu_{e}}) \delta (p_e^2) \delta
(q^2 - (p_e + p_{\nu_{e}})^2) \]
\[
\cdot \delta (E_e - p_e^0)\delta (E_{\nu_{e}} - p_{\nu_{e}}^0) \]
\begin{equation}
\cdot \sum_{X_c} \sum_{\binom{lepton}{spins}} \frac{|\langle X_c e\bar \nu_e|
H_W|
B\rangle|^2}{2m_B} (2\pi)^4 \delta^4 (p_B - (p_e + p_{\nu_{e}}) - p_{X_{c}}),
\end{equation}
where we have used formula,  $d^3 p/2p^0 = d^4 p \delta (p^2 - m^2)$.  We work
in the rest frame of the $B$ meson.  
After summing over final hadronic states $X_c$ the
only relevant angle is that between electron and neutrino.  Integrating over
the direction of neutrino gives $4\pi$  Therefore
\begin{equation}
d^3 p_e d^3 p_{\nu_{e}} = (4\pi) (2\pi) |\vec p_e|^2 d|\vec p_e| |\vec
p_{\nu_{e}}|^2 d |\vec p_{\nu_{e}}| d\cos \theta.
\end{equation}
Using $\delta (E_e - p_e^0)$ and $ \delta (E_{\nu_{e}} - p_{\nu_{e}}^0)$ to do
integrations over $p_e^0$ and $ p_\nu^0$, $\delta (p_e^2) = \delta (E_e^2 - |\vec
p_e|^2)$ to do integration
over $|\vec p_e|$, $\delta (p_{\nu_{e}}^2) = \delta (E_{\nu_{e}}^2 - |\vec
p_{\nu_{e}}|^2)$ to do
integration over $|\vec p_{\nu_{e}}|$ and $\delta (q^2 - (p_e + p_{\nu_{e}})^2)
= \delta (q^2 -
2 E_e E_{\nu_{e}} (1 - \cos \theta))$ to  perform integration over $\cos\theta$
we have
\begin{equation}
\frac{d\Gamma}{dq^2 dE_e dE_{\nu_{e}}} = \sum_{X_c}
\sum_{\binom{lepton}{spins}} \frac{|\langle X_c e\bar \nu_e | H_W
|B\rangle|^2}{8m_B} \delta^4 (p_B -
(p_e + p_{\nu_e}) - p_{X_{c}}).
\end{equation}
The square of the weak matrix element can be factored into a leptonic matrix
element and a hadronic matrix element, since leptons do not have strong
interactions
\begin{align}
&\sum_{X_c} \sum_{\binom{lepton}{spins}} \frac{|\langle X_c e\bar \nu_e| H_W|
B\rangle|^2}{2m_B} (2\pi)^3 \delta^4 (p_B - (p_e + p_{\nu_{e}}) -
p_{X_{c}})\notag \\
& = 2 G_F^2 | V_{cb}|^2 W_{\alpha\beta} L^{\alpha\beta}.
\end{align}
The lepton part $L^{\alpha\beta}$ is
\[
L^{\alpha\beta} = 4 (p_e^\alpha p_\nu^\beta + p_e^\beta p_\nu^\alpha -
g^{\alpha\beta} p_e p_\nu - i\varepsilon^{\eta\beta\lambda\alpha} p_{e\eta}
p_{\nu_e\lambda}).\]
The hadronic part is
\[
W_{\alpha\beta} = \sum_{X_c} (2\pi)^3 \delta^4 (p_B - q - p_{X_{c}})
\frac{1}{2m_B}
\langle B(p_B) | J_L^{\dagger\alpha} | X_c (p_X)\rangle\]
\begin{equation}
\cdot \langle X_c (p_X) | J_L^\beta | B (p_B)\rangle,
\end{equation}
where $J_L$ is the left handed current
\begin{equation}
J_L^\alpha = \bar c \gamma^\alpha \frac{(1 - \gamma_5)}{2} b.
\end{equation}
In the above $q = p_e + p_{\nu_{e}}$ is sum of electron
 and anti-neutrino four-momentum.
$W_{\alpha\beta}$ is a second rank tensor.  It depends on $p_B = m_B v$ and $q$
the momentum transfer to the hadronic system.  The most general tensor is
\begin{align}
W_{\alpha\beta} = &- g_{\alpha\beta} W_1 + v_\alpha v_\beta W_2 -
i\varepsilon_{\alpha\beta\mu\nu} v^\mu q^{\nu} W_3\notag \\
&+ q_\alpha q_\beta W_4 + (v_\alpha q_\beta - v_\beta q_\alpha) W_5,
\end{align}
where  $W_j = W_j (q^2, q\cdot v)$ are scalar functions.  Using this
decomposition for
$W_{\alpha\beta}$
\begin{align}
\frac{d\Gamma}{dq^2 dE_e dE_{\nu_{e}}}& = \frac{G_F^2 |V_{cb}|^2}{2\pi^3} \{W_1
q^2 + W_2(2E_e E_{\nu_{e}} - q^2/2)\notag \\
&+ W_3 q^2 (E_e - E_{\nu_{e}})\} \theta (- q^2 /4E_e + E_{\nu_{e}}),
\end{align}
where we have explicitly displayed the $\theta$-function that sets the limit on
neutrino energy.  The functions $W_4$ and $W_5$ don't contribute to the decay
rate since $q^\alpha L_{\alpha\beta} = q^\beta L_{\alpha\beta} = 0$.

The neutrino is not observed and integrating the above expression over
$E_{\nu_{e}}$ or equivalently over $v \cdot q = E_e + E_{\nu_{e}}$ gives the
measurable decay distribution $d\Gamma/dq^2 dE_e$. 
Since $q^2 = 2 E_e E_{\nu_{e}} (1 - \cos\theta)$ for fixed $E_e$ the minimum
value of $q^2$ occurs for $\cos\theta = 1$ and the 
maximum for $\cos\theta = -1$. The maximum neutrino energy and $q^2$ at fixed
electron energy are,
\[
E_{\nu_{e}}^{max} = (m_B^2 - 2m_B E_e - m_{X_{c}^{min}}^2)/(2 m_B - 4 E_e),\]
\begin{equation}
q^2 <4E_e E_{\nu_{e}}^{max} = \frac{2 E_e}{(m_B - 2E_e)} (m_B^2 - 2 m_B E_e -
m_{X_{c}^{min}}^2).
\end{equation}
The maximum electron energy is at $q^2 = 0~ ({\rm i.e.}~E_{\nu_{e}} = 0)$,
\begin{equation}
E_e^{max} = \frac{m_B^2 - m^2_{X_{c}^{min}}}{2m_B}.
\end{equation}
For a given final hadronic mass $m_{X_{c}}$, electron energy $E_e$ and $q^2$
the neutrino energy is fixed.  Using $p_X^2 = (p_B - p_e - p_{\nu_{e}})^2$
gives the neutrino energy as a function of $E_e, q^2$ and $m_{X_{c}}$,
\begin{equation}
E_{\nu_{e}} = \frac{m_B^2 + q^2 - m_{X_c}^2}{2m_B} - E_e.
\end{equation}
So integrating over $E_{\nu_{e}}$ at fixed $q^2$ and $E_e$ is equivalent to
averaging
over a range of final hadronic invariant masses.  For values of $q^2, E_e$ near
the upper boundary of the $(E_e,q^2)$ Dalitz plot
\begin{equation}
(m_B - 2 E_e)q^2 - 2E_e (m_B^2 - 2m_B E_e - m_{X_{c}^{min}}^2) = 0,
\end{equation}
only final hadronic invariant masses near $m_{X_{c}^{min}}=m_D$ get averaged over
in the integration over $E_{\nu_{e}}$.

The hadronic tensor $W_{\alpha\beta}$ parametrizes all the strong interaction
physics relevant for inclusive semileptonic $B$ decay.  It can be
related to the discontinuity of a time ordered product of currents across a
cut.  Consider
\begin{equation}
T_{\alpha\beta} = -i\int d^4 xe^{-iq \cdot x} \frac{\langle B|
T\{J_{L\alpha}^\dagger
(x) J_{L\beta} (0)\} |B\rangle}{2m_B}.
\end{equation}
Inserting a complete set of states between the time orderings
\begin{align}
T_{\alpha\beta} &= - i \int d^4 x e^{-iq\cdot x}\Bigg\{\theta (x^0)
\sum_{X_{c}} \frac{\langle B| J_{L\alpha}^{\dagger}(x) | X_c\rangle \langle X_c |
J_{L\beta}(0)|B\rangle}{2m_B}\notag \\
&+ \theta (-x^0) \sum_{X_{\bar cbb}} \frac{\langle B | J_{L\beta} (0) | X_{\bar
c bb}\rangle \langle X_{\bar c bb}| J_{L\alpha}^{\dagger} (x) | B\rangle}{2m_B}\Bigg\},\notag \\
\end{align}
and performing the space-time integration gives
\begin{align}
T_{\alpha\beta} &= \sum_{X_{c}} \frac{\langle B|J_{L\alpha}^\dagger(0) | X_c
\rangle \langle X_c | J_{L\beta}(0)
|B\rangle}{2m_B (m_B - E_{X_{c}} - q^0 + i\varepsilon)} (2\pi)^3 \delta^3
(\vec p_{X} + \vec q)\notag \\
& - \sum_{X_{\bar c bb}} \frac{\langle B|J_{L\beta}(0) | X_{\bar c bb} \rangle
\langle X_{\bar c bb} | J_{L\alpha}^\dagger(0) |B\rangle}{2m_B (E_{X_{\bar c bb}}
- m_B - q^0 - i\varepsilon)} (2\pi)^3 \delta^3 (\vec p_{X} - \vec q).
\end{align}
At fixed $\vec q$ the time ordered product $T_{\alpha\beta}$ has cuts in the
complex $q^0$ plane along the real axis.  One cut is in the region $-\infty <
q^0 < m_B - \sqrt{|\vec q|^2 + m_{X_{c}^{min}}^2}$ and other is in the region
$\sqrt{|\vec q|^2 + m_{X_{\bar c bb}^{\min}}^2} - m_B <q^0 <\infty.$
The two cuts are well separated.  The discontinuity in $T$ across its
cuts is evaluated using the formula
\[
\frac{1}{\omega + i\varepsilon} = P \frac{1}{\omega} - i\pi \delta (\omega).\]
This gives
\begin{align}
\frac{1}{\pi} Im T_{\alpha\beta} &= - \sum_{X_{c}} \frac{\langle
B|J_{L\alpha}^{\dagger}(0) |X_c
\rangle \langle X_c | J_{L\beta}(0) |B\rangle}{2m_B} (2\pi)^3 \delta^4(p_B -
p_X - q)\notag \\
&- \sum_{X_{\bar c bb}} \frac{\langle B |J_{L\beta}(0) | X_{\bar c bb} \rangle
\langle
X_{\bar c bb} | J_{L\alpha}^{\dagger}(0) |B \rangle}{2m_b} (2\pi)^3 \delta ^4 (p_B + q
- p_X).
\end{align}
The first of these terms is just $- W_{\alpha\beta}$.  For $q$ in the
region of semileptonic decay the second delta function is not satisfied and so
\begin{equation}
- \frac{1}{\pi} Im T_{\alpha\beta} = W_{\alpha\beta}.
\end{equation}
It is convenient to express $T_{\alpha\beta}$ in terms of Lorentz scalar form
factors as we did for $W$.
\begin{align}
T_{\alpha\beta} &= - g_{\alpha\beta} T_1 + v_\alpha v_\beta T_2 -
i\varepsilon_{\alpha\beta\mu\nu} v^\mu q^\nu T_3\notag \\
&+ q_\alpha q_\beta T_4 + (v_\alpha q_\beta + v_\beta q_\alpha) T_5.
\end{align}
The $T_j$'s are also functions of $q^2$ and $q \cdot v$.  One can study 
the $T_j$'s
in the complex $q \cdot v$ plane at fixed $q^2$.  This is the Lorentz invariant
way of phrasing the analytic structure.  For the cut associated with physical
hadronic states containing a $c$ quark $(p_B - q) - p_X = 0$, and so $m_B^2 - 2
v \cdot q m_B + q^2 = m^2_{X_c}$.  Therefore
\begin{equation}
 v \cdot q <
\frac{(m_B^2 + q^2 - m_{X_{c}^{\min}}^2)}{2m_B}.
\end{equation}
On the other hand cut corresponding to states with $\bar c bb$ quark content,
$(p_B + q) - p_X = 0$, implying that $m_B^2 + 2 v \cdot q m_B + q^2 =
m^2_{X_{\bar c bb}}$.  In this case
\begin{equation}
v\cdot q > \frac{(m_{X_{\bar c bb}^{min}}^2 - m_B^2 - q^2)}{2m_B}.
\end{equation}

\begin{figure}[t]
\centerline{\unitlength=1cm \epsfysize=5cm \epsfbox{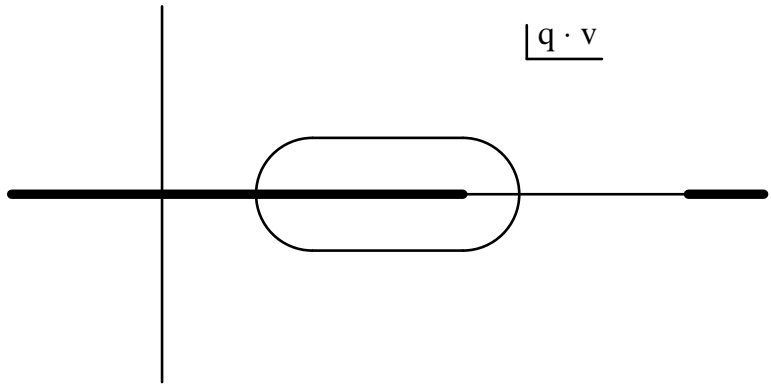}}
\vspace{\baselineskip}
\caption{Contour for recovering integrals of the $W_j$'s from those of the
$T_j$'s.}
\end{figure}

The differential decay rate $d\Gamma/d^2 q E_e$ is obtained from $d\Gamma/dq^2
d E_e d
v \cdot q$ by integration over $v \cdot q$.  Integrals over $v \cdot q$ of the
structure functions $W_j (q^2, v \cdot q)$ are related to integrals of $T_j$
over contour shown in Fig. (2).

The structure functions $T_j$ can be expressed in terms of matrix elements of
local
operators using the operator product expansion (OPE) to simplify the time
ordered product~\cite{shifmanvoloshin,chay,bigi,manohar}.
\begin{equation}
-i \int d^4 xe^{-i \cdot q \cdot x} 
T \{J^\dagger_{L\alpha} (x) J_{L\beta} (0)\}.
\end{equation}
The $B$-meson matrix element of the above time ordered product is
$T_{\alpha\beta}$.  The coefficients of the
operators that occur can be reliably computed in perturbative QCD in a region
that is far away from the cuts.  These operators involve quark and gluon
fields. They can be expanded in powers of $1/m_b$ by making a transition to
HQET.

Note that the contour in Fig. (2) necessarily touches the cut. However this
is in a region that is far from the upper boundary of the 
$(E_e,q^2)$ Dalitz plot and is a place where the kinematic restrictions are on
the production of high mass states (i.e., $m_{X_c}>>m_D$). 
In this region the OPE
is expected to give inclusive differential decay rates that are valid
locally (i.e, without averaging over final state hadronic masses) up to
small corrections. If the contour pinched the cut near its endpoint at the
right, there would be very large corrections since this corresponds to a
kinematic region near the upper boundary of the $(E_e,q^2)$ 
Dalitz plot where the rate is
dominated by the production of low mass final hadronic states with masses
near $m_D$.

At lowest order in perturbation theory the matrix element of this time
ordered product between $b$-quark states with residual momentum $k$ is
\begin{equation}
\frac{1}{(m_b v - q + k)^2 - m_c^2 + i\varepsilon} \bar u \gamma_\alpha
\frac{(1-\gamma_5)}{2} (m_b \vslash - \qslash + \kslash) \gamma_\beta
\frac{(1-\gamma_5)}{2} u.
\end{equation}
From eq. (11.23) it is evident that (without including perturbative corrections)
the discontinuity across the cut in Fig. (2) is concentrated at its
endpoint. This is another reason why it is not a problem to have the contour
touching the cut far from this point.

\section{ $b$-quark Decay}

The leading contribution to the decay rate comes from the order $k^0$ terms in
eq. (11.23).  Reducing the product of 3 gamma matrices to a sum of single gamma
matrices it becomes
\[
\frac{1}{\Delta_0} \bar u \{(m_b v - q)_\alpha \gamma_\beta + (m_b v - q)_\beta
\gamma_\alpha - (m_b \vslash - \qslash) g_{\alpha\beta}\]
\begin{equation}
- i \varepsilon_{\alpha\beta\lambda n} (m_b v - q)^\lambda \gamma^n \}\frac{(1
-
\gamma_5)}{2} u,
\end{equation}
where
\begin{equation}
\Delta_0 = (m_b v - q)^2 - m_c^2 + i\varepsilon.
\end{equation}
Matrix elements of $\bar b \gamma_\lambda b$ and $\bar b \gamma_\lambda\gamma_5
b$ between $b$-quark states are $\bar u \gamma_\lambda u$ and $\bar u
\gamma_\lambda\gamma_5 u$.  Hence the leading term in the OPE is obtained by
replacing $u$ above with $b$-quark field.  To get the $T_j$'s we need  the $B$
meson matrix elements
\begin{align}
\langle B(p_B) | \bar b\gamma_\lambda b|B(p_B)\rangle &= 2 p_{B\lambda} = 2
m_B v_\lambda, \notag \\
\langle B(p_B)|\bar b \gamma_\lambda \gamma_5 b|B(p_B) \rangle &= 0.
\end{align}
Note that these are exact.  In this case there is no need to make a
transition to HQET to determine the $m_b$ dependence of the matrix element.
The first of eqs. (12.3) follows from
b-quark number conservation and the second from parity invariance.  The
resulting structure functions are easily found.  For example
\[
T_1 = \frac{1}{2\Delta_0} (m_b - q \cdot v),\]
which implies that
\begin{align}
W_1 &= - \frac{1}{\pi} Im T_1 = \frac{(m_b - q \cdot v)}{2} \delta ((m_b v -
q)^2 - m_c^2)\notag \\
&= \frac{(m_b - q \cdot v)}{4 m_b} \delta \left(v \cdot q - \left(\frac{q^2 +
m_b^2 -
m_c^2}{2m_b}\right)\right).
\end{align}
The leading term in the OPE gives functions $W_j$ that correspond to free
$b$-quark decay.  Consequently we have derived  the
usual result that at leading order in
$\Lambda_{QCD}/m_b$ the $B$ meson semileptonic decay rate is equal to the
$b$-quark semileptonic decay rate.

\section{The Chay--Georgi--Grinstein Theorem}

At linear order in $k$ the $b$-quark matrix element of the time ordered product
of currents gives
\[
\frac{1}{\Delta_0} \bar u \{k_\alpha \gamma_\beta + k_\beta \gamma_\alpha -
g_{\alpha\beta} \kslash - i \varepsilon_{\alpha\beta\lambda n} k^\lambda
\gamma^n \} \frac{(1 - \gamma_5)}{2} u\]
\begin{align}
&- \frac{2k \cdot (m_b v - q)}{\Delta_0^2} \bar u \{(m_b v - q)_\alpha
\gamma_\beta + (m_b v - q)_\beta \gamma_\alpha\notag \\
&- (m_b\vslash - \qslash) g_{\alpha\beta} - i \varepsilon_{\alpha\beta\lambda
\eta} (m_b v - q)^\lambda \gamma^\eta\}\frac{(1-\gamma_5)}{2} u.
\end{align}
After converting to HQET $b$-quark fields these produce terms in the operator
product expansion corresponding to the operators $\bar b_v \gamma_\lambda
iD_\tau b_v$ and $\bar b_v
\gamma_\lambda \gamma_5 i D_\tau b_v$.  The second of these has zero $B$ meson
matrix
element by parity.  For the first use $\bar b_v \gamma_\lambda i D_\tau
b_v = v_\lambda \bar b_v i D_\tau b_v$.  The general form for the $B$ meson
matrix element of this operator is
\begin{equation}
\langle B(v)| \bar b_v i D_\tau b_v |B(v)\rangle = X v_{\tau}.
\end{equation}
Contracting with $v^\tau$ implies
\begin{equation}
X = \langle B(v)| \bar b_v iv \cdot D b_v |B(v)\rangle = 0,
\end{equation}
by the equations of motion.  So there are no $\Lambda_{QCD}/m_b$ corrections to
$b$-quark decay picture for inclusive $B$ meson decay ~\cite{chay}.

\section{Higher Order Corrections to  the Inclusive B Decay Rate.}

We have seen that up to corrections suppressed by $\alpha_s (m_b)$ and
$\Lambda^2_{QCD}/m_b^2$ the inclusive semileptonic $B$ meson decay rate is
equal to the free $b$-quark decay rate.

The perturbative corrections can be included by calculating the perturbative
corrections to the $b$-quark decay rate.  This has been done to order
$\alpha_s^2 (m_b)$~\cite{savage,cz}.

At order $\Lambda_{QCD}^2/m_b^2$ dimension 5 operators occur in the OPE.  They
are the same operators that occur in the $B$ meson mass formula.  At this order
the nonperturbative corrections to the inclusive semileptonic $B$ decay rate
involve $\lambda_1$ and $\lambda_2$ ~\cite{bigi,manohar}. Similar results hold for the inclusive
nonleptonic $B$ decay rate.

\section{NRQCD}

HQET is not the appropriate effective field theory for systems with more than
one heavy quark.  In HQET the heavy quark kinetic energy is neglected.  It
occurs as a small $1/m_Q$ correction.  At short distances the static potential
between heavy quarks is determined by one gluon exchange and is a Coulomb
potential.  For a $Q\bar Q$ pair in a color singlet it is an attractive
potential
and the heavy quark kinetic energy is needed to stabilize a $Q\bar Q$ meson.
For $Q\bar Q$ hadrons (i.e., quarkonia) the kinetic energy plays a very
important role and it cannot be treated as a perturbation.

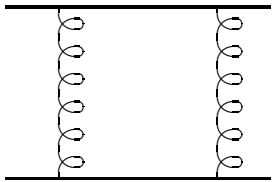
\begin{figure}[h]
\begin{center}
\begin{picture}(10000,10000)
\THICKLINES
\drawline\fermion[\E\REG](0,0)[10000]
\THINLINES
\drawline\gluon[\N\REG](2000,0)[6]
\THICKLINES
\drawline\fermion[\E\REG](0,\gluonbacky)[10000]
\THINLINES
\drawline\gluon[\N\REG](8000,0)[6]
\end{picture}
\end{center}
\vskip 0.3in
\caption{ One loop QQ scattering in HQET.}
\end{figure}

In fact the problem is more general than this.  Consider, for example, trying
to calculate low energy $QQ$ scattering in the center of mass frame
in HQET.  Setting
$v = v_r = (1,{\bf 0})$ for each heavy quark ( intitial residual momenta
are $k_{\pm}=(0,\pm {\bf k})$ and final residual momenta are $k_{\pm}'=(0,\pm {\bf k'})$) the one loop Feynman diagram in
Fig. (3) gives rise to a loop integral
\begin{equation}
\int \frac{d^n q}{(2\pi)^n} \frac{i}{(q^0 + i\varepsilon)} \frac{i}{(-q^0 +
i\varepsilon)}
 \frac{i}{((q+k_+)^2+i\varepsilon)}\frac{i}{((q+k_+')^2+i\varepsilon)}.
\end{equation}
The $q^0$ integration is ill defined because it has poles above and below the
real axis at $q^0 = \pm i\varepsilon$.  This problem is cured by not treating
the heavy
quark kinetic energy as a perturbation but rather including it in the leading
order terms.

Properties of quarkonia are usually predicted as a power series in $v/c$ where
$v$ is the magnitude of the relative $Q\bar Q$ velocity and $c$ is the speed of
light.  For these systems the appropriate limit of QCD to examine is the $c
\rightarrow \infty$ limit, where the QCD Lagrangian becomes an effective field
theory called non-relativistic quantum chromodynamics (NRQCD)~\cite{bodwin}.
  For finite $c$ there are corrections
suppressed by powers of $1/c$.  In particle physics we usually set Planck's
constant divided by $2\pi$ and the speed of light to unity. 
 Making the factors of $c$ explicit the QCD Lagrangian density is
\begin{equation}
{\cal L}_{QCD} = - \frac{1}{4} G_{\mu\nu}^B G^{B \mu\nu} - c \bar Q (i\Dslash
- m_{Q}c) Q.
\end{equation}
In the above the $0$ component of a partial derivative is
\begin{equation}
\partial_0 = \frac{1}{c} \frac{\partial}{\partial t},
\end{equation}
and $D$ is the covariant derivative
\begin{equation}
D_\mu = \partial_\mu + \frac{ig}{c} A_\mu^B T^B.
\end{equation}
The gluon field strength tensor $G_{\mu\nu}^B$ is defined in the usual way
except that $g \rightarrow g/c$.

All dimensionful
quantities can be expressed in units of length $[x]$ and time $[t]$.  (i.e.,
$[E] \sim 1/[t]$ and $[p] \sim 1/[x])$.  The Lagrangian $L = \int d^3 x {\cal
L}$ has units of $1/[t]$ since the action $S = \int {\cal L}dt$ is
dimensionless.  It
is straightforward to deduce that the gluon field has units $[A] \sim
1/\sqrt{[x] [t]}$ and the strong coupling $g \sim \sqrt{[x]/[t]}$.  The fermion
field has units $[\psi] \sim 1/[x]^{3/2}$ while its mass has units $[m_Q] \sim
[t]/[x]^2$.  With these units $m_Q c^2$ has dimensions of energy and the strong
fine structure constant $\alpha = g^2/4\pi c$ is dimensionless.

For the fermion field $Q$ the transition from QCD to NRQCD follows the
derivation of HQET.  It is rewritten as
\begin{equation}
Q = e^{-im_{Q} c^{2} t} \left[1 +\frac{i\Dslash_\perp}{m_{Q}c} + \ldots\right]
\binom{\psi}{0},
\end{equation}
where $\psi$ is a two-component Pauli spinor and the $\perp$ is with
respect to the rest frame four-velocity $v_r=(1,{\bf 0})$.  Using this, 
the part of the QCD
Lagrange density involving $Q$ becomes
\begin{equation}
{\cal L}_\psi = \psi^\dagger \left[i\left(\frac{\partial}{\partial t} + ig
A_0^B
T^B\right) + \frac{\bfnabla^2}{2m_Q}\right] \psi +\ldots,
\end{equation}
where the ellipses denote terms suppressed by powers of $1/c$.  Note that the
heavy quark kinetic energy is now leading.  The replacement $g
\rightarrow g/c$ was necessary to have a sensible $c \rightarrow
\infty$ limit.

Among the terms suppressed by a single power of $1/c$ is the gauge completion
of the kinetic energy
\begin{equation}
{\cal L}_{int} = \frac{ig}{m_{Q}c} {\bf A}^B [\psi^\dagger T^B \bfnabla \psi -
(\bfnabla \psi)^\dagger T^B \psi].
\end{equation}
There is also a $1/c$ term involving the color magnetic field ${\bf B}_c =
\bfnabla \times {\bf A}$.

It is convenient to work in Coulomb gauge, $\bfnabla \cdot {\bf A}^B = 0$.
Then the part of action that involves the gluon field strength tensor and is
quadratic in the fields simplifies to
\[
-\frac{1}{4} \int d^3 x G_{\mu\nu}^B G^{B\mu\nu} \rightarrow  \frac{1}{2} \int
d^3 x G_{0i}^B G_{0i}^B - \frac{1}{4} \int d^3 x G_{ij}^B G_{ij}^B\]
\begin{equation}
= \frac{1}{2} \int d^3 x \left( (\partial_i A_0^B)^2 + (\partial_0 A_i^B)^2 -
(\partial_i A_j^B)^2 \right) + ~{\rm nonabelian~terms}.
\end{equation}
The $0$ component of the gauge field does not propagate, and its effects on the
heavy quarks is reproduced by an instantaneous Coulomb potential.
\begin{equation}
V_c ({\bf x, y}) = \frac{g^2}{4\pi} \frac{1}{|{\bf x} - {\bf y}|}.
\end{equation}
In $Q\bar Q$ systems the $0$ component of the gluon field typically has a four-
momentum of order $(E, {\bfp}) \sim (m_Q v^2, m_Q v)$.   It is far off-shell
and can be integrated out of the theory giving rise to the Coulomb potential
above.  The
transverse gluons ${\bf A}$ have both potential modes where $(E, {\bfp}) \sim
(m_Q v^2, m_Q v)$ and propagating modes, where $(E, {\bf p}) \sim (m_Q v^2, m_Q
v^2 /c)$, that are important~\cite{labelle}.  The potential modes do not
propagate and as was done for $A_0$ the Lagrangian can be written without these
fields explicitly appearing.
However the propagating modes must appear explicitly.  For these modes the
gluon field is rewritten in terms a scaled spatial component ${\bfy} =
{\bfx}/c$ so that the $(1/c^2) \partial^2/\partial t^2$ part of the kinetic
term is as important as the $\bfnabla^2$ part of the kinetic
term~\cite{grinstein1},
\begin{equation}
{\bf A}^B({\bf x},t) = \frac{1}{\sqrt{c}} {\tilde{{\bf A}}}^B ({\bf x}/c,t).
\end{equation}
Using this the QCD Lagrangian becomes, for large $c$,
\begin{equation}
L_{QCD} = L_{NRQCD} + \ldots,
\end{equation}
where the ellipsis denote terms suppressed by powers of $1/c$ and
\begin{align}
L_{NRQCD} &= \int d^3 x \psi^\dagger \left( i \frac{\partial}{\partial t} +
\frac{\bfnabla^2}{2m_Q}\right)\psi\notag \\
&- \!\int\!\! d^3 x_1 d^3 x_2 \psi^\dagger\! ({\bfx}_1,\! t) T^A \psi
({\bfx}_1, t) V_c ({\bfx}_1, {\bfx}_2) \psi^\dagger ({\bfx}_2, t) T^A \psi
({\bfx}_2, t)\notag \\
&+ \frac{1}{2} \int d^3 y \left(\left(\frac{\partial}{\partial t} \tilde{A}_i^B
({\bfy},
t)\right)^2 - \left(\frac{\partial}{\partial y^i} \tilde{A}_j^B
({\bfy},t)\right)^2\right),
\end{align}
with ${\bfy} = {\bfx}/c$.

At leading order in the $1/c$ expansion the radiation gluons do not interact
with the heavy
quarks.  Also the nonabelian terms in $G_{\mu\nu}^B G^{B\mu\nu}$ are
suppressed.  The leading interaction of the propagating gluons with the heavy
quark is
\begin{equation}
{\cal L}_{int} = \frac{ig}{m_{Q}c^{3/2}} {\tilde{{\bfA}}}^B ({\bf 0},t)
[\psi^\dagger T^B
\bfnabla \psi - (\bfnabla \psi)^\dagger T^B \psi].
\end{equation}
Note that the term involving the propagating gluon color magnetic field is less
important because $\tilde{{{\bf B}}}_c = \vec\nabla \times \tilde{{\bfA}}
\left({\bfx}/c, t \right) = 0 + \frac{1}{c} \bfnabla \times
\tilde{{\bfA}} (0,t) + \ldots$ is suppressed by $1/c$.  This derivation of
NRQCD has followed closely that in Ref.~\cite{grinstein1}.  (For a somewhat
different scaling of fields see ~\cite{luke3}.)

The NRQCD Lagrangian does not have a heavy quark flavor symmetry because $m_Q$
explicitly appears in the kinetic energy term.  However it does have the spin
symmetry.  Furthermore, the leading interaction of the propagating gluons with
the heavy quarks also respects the heavy quark spin symmetry.

In some respects the Feynman rules for this theory are a little unusual.
Coulomb exchange is denoted by a dashed line corresponding to a propagator
$-i/{\bfk}^2 $ and a vertex $-igT^A$.  The heavy quark,
$\psi$, propagator is
\begin{equation}
\frac{i}{k^0 - \frac{{\bfk}^2}{2m_Q} + i\varepsilon}.
\end{equation}
The transverse radiation gluons are denoted by the usual wavy line. Their
propagator is
\begin{equation}
i \frac{(\delta_{ij} - k_i k_j/{\bfk}^2)}{(k^{0})^2 - {\bfk}^2 + i\varepsilon},
\end{equation}
and their interaction with heavy quarks is determined by eq. (15.13).  It
gives rise to Feynman rules where energy is conserved at the vertices but not
the three-momentum.  The transverse radiation gluons do not transfer any three
-momentum to the heavy quarks.

Anti-quarks can be added in the same way as in HQET.  For these one writes
\begin{equation}
Q = e^{im_{Q}c^{2} t} \left[1 + \frac{i \Dslash_\perp}{2m_Q c} + \ldots\right]
\binom{0}{\chi},
\end{equation}
and their coupling is similar to that of the quarks.  The field $\chi$ creates
a heavy anti-quark while $\psi$ destroys a heavy quark.

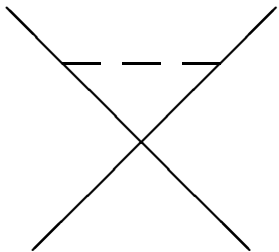
\begin{figure}[h]
\begin{center}
\begin{picture}(10000,5000)
\drawline\scalar[\E\REG](0,0)[3]
\THICKLINES
\drawline\fermion[\NW\REG](\scalarfrontx,\scalarfronty)[3000]
\drawline\fermion[\SE\REG](\scalarfrontx,\scalarfronty)[10000]
\drawline\fermion[\NE\REG](\scalarbackx,\scalarbacky)[3000]
\drawline\fermion[\SW\REG](\scalarbackx,\scalarbacky)[10000]
\end{picture}
\end{center}
\vskip 0.9in
\caption{ Contribution to matrix element in Eq. (15.17). }
\end{figure} 

The formalism developed above helps us understand the size of matrix elements
in the $v/c$ expansion.  For the first example consider
\begin{equation}
\langle~^3S_1| (\psi^\dagger \sigma^i \chi) (\chi^\dagger \sigma^i
\psi)|~^3S_1\rangle.
\end{equation}
Here $|~^3S_1\rangle$ denotes a $Q\bar Q$ state at rest with $~^3S_1$ quantum
numbers (with the usual $^{(2S + 1)}L_J$ spectroscopic notation).  For $c\bar
c$ quarkonia the lowest mass state with these quantum numbers is the $\psi$ and
for $b\bar b$ it is the $\Upsilon$.  At leading order it is only the Coulomb
potential gluons that are relevant.  Feynman diagrams where a potential gluon
is exchanged between the
quark-anti-quark pairs in eq. (15.17) vanish.  Fig. (4) is such a diagram.  The loop
integration
is of the form,
\begin{equation}
\int \frac{d^n k}{(2\pi)^n} \frac{1}{(E + k^0) - \frac{({\bfk} +
{\bfp})^2}{2m_Q} + i\varepsilon} \cdot \frac{1}{(E' + k^0) - \frac{({\bfk} +
{\bfp}')^2}{2m_Q} + i\varepsilon} \cdot \frac{i}{{\bfk}^2},
\end{equation}
which is zero because all the poles in the $k^0$ integration are on the same
side of the real axis.  This generalizes to all orders in perturbation theory
and consequently the matrix element factorizes.
\begin{equation}
\langle ~^3S_1 |(\psi^\dagger \sigma^i \chi) (\chi^\dagger \sigma^i
\psi)|~^3S_1\rangle = \langle ~^3S_1 | (\psi^\dagger \sigma^i \chi)|0\rangle
\langle 0| (\chi^\dagger \sigma^i \psi)|~^3S_1\rangle.
\end{equation}
Corrections to factorization come from radiation gluons and so are suppressed
by powers of $v/c$.

 Next consider a color-octet matrix element,
\begin{equation}\label{*}
\langle~^3S_1|(\psi^\dagger T^A \sigma^i \chi) (\chi^\dagger \sigma^i T^A
\psi)|~^3S_1\rangle.
\end{equation}
Since $|~^3S_1\rangle$ is a color singlet at least one radiation gluon must be
exchanged between the quark anti-quark pairs.  Such a diagram is shown in Fig.
(5).  
\begin{figure}[h]
\begin{center}
\begin{picture}(10000,5000)
\drawline\gluon[\E\FLIPPED](0,0)[5]
\THICKLINES
\drawline\fermion[\NW\REG](\gluonfrontx,\gluonfronty)[3000]
\drawline\fermion[\SE\REG](\gluonfrontx,\gluonfronty)[10000]
\drawline\fermion[\NE\REG](\gluonbackx,\gluonbacky)[3000]
\drawline\fermion[\SW\REG](\gluonbackx,\gluonbacky)[10000]
\end{picture}
\end{center}
\vskip 0.9in
\caption{ Contribution to matrix element in Eq. (15.20). }
\end{figure}
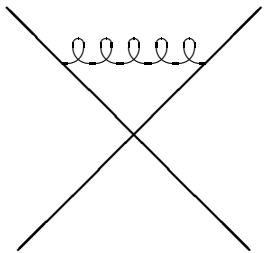 
But the
coupling of the transverse gluon is proportional to ${\bfp}$ and averaging with
the $S$-wave, $|~^3S_1\rangle$ momentum space wavefunction, causes this diagram
to
vanish.  Two transverse gluon exchanges are needed to get a non-zero amplitude
and so this matrix element occurs at order $(v/c)^6$.  (Recall ${\cal L}_{int}$
is proportional to $1/c^{3/2}$.)

This power counting is correct for very heavy quarks where $m_Q v^2/c\gg
\Lambda_{QCD}$.   For $\bar bb$ and $\bar cc$ systems this inequality is not
satisfied.  Often for these systems one uses a modified power counting which is
appropriate for $m_Q v^2/c \sim \Lambda_{QCD}$.  In this situation the coupling
of the radiation gluons should be nonperturbative with $\alpha$ of order unity
and $g \sim \sqrt{c}$.  Then ${\cal L}_{int}$ is suppressed
by only $1/c$ and the leading contribution to the color-octet matrix element in
eq.~(\ref{*}) is suppressed by $(v/c)^4$ instead of $(v/c)^6$.

It is easy to see that with $g \sim \sqrt{c}$ for radiation gluons their
nonabelian self couplings are
not suppressed.  For example, the three gluon coupling
\begin{align}
{\cal L}_{3gluon} &\sim \frac{g}{c} \int d^3 x
\left(\frac{1}{\sqrt{c}}\right)^3
\{\bfnabla \tilde{{\bfA}} ({\bfx}/c, t)) \tilde{{\bfA}}
({\bfx}/c, t) \tilde{{\bfA}} ({\bfx}/c, t)\}\notag \\
&\sim \frac{g}{\sqrt{c}} \int d^3 y ({\bfnabla}_{\bfy} \tilde{{\bfA}}
({\bfy},t)) \tilde{{\bfA}} ({\bfy},t)\tilde{{\bfA}} ({\bfy},t),
\end{align}
is of order unity if $g \sim \sqrt{c}$.  Since the quark coupling to
the radiation gluons is suppressed, even with $m_Q v/c^2 \sim
\Lambda_{QCD}$, the leading order quarkonia states are 
still determined by the Coulomb
potential. (Note that the strong coupling for potential gluons retains
the usual scaling $g \sim 1 $.)
 For the rest of this section this modified power counting
appropriate to $m_Q v^2/c \sim \Lambda_{QCD}$ is used.  (It would be useful to
have a more systematic understanding of this modification of the power
counting.)

Next consider the matrix element
\begin{equation}
\langle ~^3S_1| (\psi^\dagger \frac{\vec D}{m_Qc} T^A \sigma^i \chi)
(\chi^\dagger \frac{\vec D}{m_Q c} T^A\sigma^i \psi)|^3S_1\rangle.
\end{equation}
The factors of $m_Qc$ in the denominator are inserted so that this
operator has the same dimension as those occuring in eqs. (15.20) and (15.17).
Now only a single radiation gluon is needed between the quark pairs.  This
matrix element is also suppressed compared to that in eq. (15.17) by $(v/c)^4$.
Two of the factors of $1/c$
coming from the radiation gluon vertex and two from the explicit
$c$'s in the operator.  

Radiation gluons have a novel coupling to heavy quarks.   Usually in the
derivation of the Feynman rules there is an integral $\int d^4 k/(2\pi)^4$ for
each propagator and a $(2\pi)^4\delta^4 (k_i - k_f)$ for each vertex.  The
result is that there is one $\int d^4 k/(2\pi)^4$ integration for each loop.
But the leading vertex for the $\psi(p) \rightarrow \psi (p') \tilde{{\bfA}}
(k)$ radiation gluon heavy quark interaction in ${\cal L}_{int}$ contains the
delta function $(2\pi)^4 \delta (E - E' - k^0)\delta^3 ({\bfp} -  {\bfp}')$.
At leading order in $v/c$ the radiation gluons do not transfer three-momentum.
One consequence of this is that the matrix element
\begin{equation}
\langle~^3S_1|(\psi^\dagger \frac{\vec D}{m_Q c} T^A \sigma^i
\chi) \frac{D_{i_{1}}}{m_Q c} \ldots \frac{D_{i_{m}}}{m_Q c} (\chi^\dagger
\frac{\vec D}{m_Q c} T^A \sigma^i \psi)|~^3S_1\rangle,
\end{equation}
is suppressed compared with
\begin{equation}
\langle ~^3S_1 |(\psi^\dagger \frac{\vec D}{m_Q c} T^A \sigma^i
\chi) (\chi^\dagger \frac{\vec D}{m_Q c} \frac{D_{i_{1}}}{m_Q c} \ldots
\frac{D_{i_{m}}}{m_Q c} T^A \sigma^i \psi)| ~^3S_1\rangle.
\end{equation}
For the former matrix element the Feynman diagram in Fig. (5) 
contains the factor $(p' -
p)^{i_{1}} \ldots (p' - p)^{i_{m}}$ from the spatial derivatives on the $Q\bar
Q$
quark pair.  In the rest frame of the $^3S_{1}$ state this vanishes unless
one expands $\tilde{{\bfA}}({\bfx}/c, t)$ in ${\cal L}_{int}$
to $m$ powers of $1/c$~\cite{grinstein2}.  The $m$'th term in the expansion of
$\tilde{{\bfA}}$ gives a factor
$x^{i_{1}}\ldots x^{i_{m}}/c^m \partial_{i_{1}} \ldots
\partial_{i_{m}} \tilde{{\bfA}} (0,t)$ producing in the Feynman diagram the
factor
$(k^{i_{1}} \ldots k^{i_{m}})/c^m \partial_{i_{1}} \ldots$ $\partial_{i_{m}}
\delta^3 ({\bfp}' - {\bfp})$.  (The factors of $x_i$ become
derivatives $\partial_i$ acting on the delta function.) 
The Feynman integration over ${\bfp}'$ is done
by integrating by parts $m$ times
\[
\int d^3 p' (p' - p)_{j_{1}} \ldots (p' - p)_{j_{m}} \partial_{i_{1}} \ldots
\partial_{i_{m}} \delta^3 ({\bfp}' - {\bfp}) \]
\begin{equation}
= (\delta_{i_{1}j_{1}} \ldots \delta_{i_{m}j_{m}} + perms).
\end{equation}
The matrix element with the $m$ derivatives  acting on the $\chi^\dagger \psi$
pair is
suppressed by $(v/c)^m$ compared with the matrix element where the $m$
derivatives act between the quarks in the $\chi^\dagger \psi$ pair.

One of the most important recent developments in quarkonia physics is the
realization that octet matrix elements suppressed by powers of $v/c$ are
sometimes more important than the analogous color singlet matrix element
because they have coefficients that
are enhanced by factors of $1/\alpha_s (m_Q)$.  A simple example
of this occurs for production of quarkonia (with $^3S_1$ quantum numbers) at large momentum transverse to the beam in
$p\bar p$ scattering~\cite{braaten}.  At the present time this has only been
observed for the case $Q = c$ corresponding to $\psi$ or $\psi'$ production.

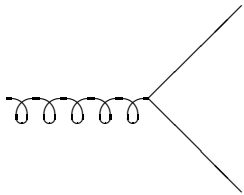
\begin{figure}[h]
\begin{center}
\begin{picture}(10000,3000)
\THINLINES
  \drawline\gluon[\E\REG](1000,0)[5]
    \drawline\fermion[\NE\REG](\gluonbackx,\gluonbacky)[5000]
  \drawline\fermion[\SE\REG](\gluonbackx,\gluonbacky)[5000]
\end{picture}
\end{center} 
\vskip 0.3in
\caption{ Color-octet quark anti-quark production from fragmenting gluon. }
\end{figure}

The production of quarkonia at large transverse momentum
 is dominated by gluon fragmentation.   The underlying process is a virtual
gluon $k_g^2 \simeq 4m_c^2$ converting into a $c\bar c$ pair in a $^3S_1$,
color
octet configuration.  The color octet $c\bar c$ pair omits
radiation gluons producing a $\psi X$ final state.  The differential cross
sections is written as
\begin{equation}\label{**}
d\sigma = \int dz d\sigma_{gg\rightarrow gg} (k_g = p_\psi/z)
D_{g\rightarrow \psi} (z) ,
\end{equation}
where the $\psi$ carries a fraction $z$ of the gluons
momentum.  The square of the diagram in Fig. (6)
gives a gluon fragmentation function that
is proportional to $\delta (1-z)$.  Explicitly
\begin{equation}
D_{g \rightarrow \psi} (z) \propto g^2(2m_c) \left(\frac{1}{4m_c^2}\right)^2
\delta (1 - z) \langle 0^{(^3S_{1})}\rangle,
\end{equation}
where $\langle 0^{(^3S_{1})}\rangle$ is proportianal to
the fragmentation probability.  In perturbation theory it can be written as,
\begin{equation}
\langle 0^{(^3S_{1})}\rangle = \sum_X \langle 0 | (\chi_c^\dagger T^A
\sigma^i \psi_c)|\psi + X\rangle \cdot \langle \psi + X | (\psi_c^\dagger T^A
\sigma^i\chi_c)|0\rangle.
\end{equation}
The hadronic matrix element above is evaluated in the $\psi$ rest
frame and a sum over $\psi$ polarizations is understood.  The
fragmentation function is proportional to $\delta (1 - z)$ since the radiation
gluons carry away a fraction of the $\psi$'s momentum suppressed by $(v/c)^2$.
The factor of $(1/4m_c^2)^2$ in eq. (15.27) comes from squaring the gluon
propagator.   To
write the cross section in the form above, the phase space factor $d^3
p_{\psi}/2p_{\psi}^0$ is set equal to $d^3 k_g/2k_g^0$ since for a
high energy process the $\psi$ mass can be neglected in
$p_{\psi}^0$.  The radiation
gluons hadronize into a set of pions accompanying the $\psi$ that
carry a fraction of order $v^2/c^2$ of the $\psi$'s energy.

The
fragmenting gluon is almost on shell.  An on shell gluon has only transverse
polarizations and so the color octet $c\bar c$ pair produced by the fragmenting
gluon is also
transversely aligned.  Since the leading couplings of the radiation gluons
to the heavy quark preserve spin symmetry this transverse alignment is
transferred to the $\psi$ in the final state~\cite{chowise}.  (There are $v/c$ ~\cite{beneke1} and perturbative $\alpha_s$
corrections~\cite{beneke2} that reduce this alignment somewhat.)

There is a color singlet contribution to the $g \rightarrow \psi$
fragmentation function.  It is calculated from Feynman diagrams like Fig (6),
but with two hard gluons radiated off the quark legs to make the heavy quark 
anti-quark pair in a color singlet configuration.  This contribution to the fragmentation function  is enhanced to
$(c/v)^4$ compared to that in eq. (15.27).  However, it is suppressed by
$(\alpha_s (2m_c)/\pi)^2$ because of the two additional hard gluons.

\section{Conclusions}

Over the last decade there has been remarkable progress in our ability to
predict properties of hadrons containing a single heavy quark and properties of
hadrons containing a heavy quark anti-quark pair (i.e., quarkonia).

The application of heavy quark symmetry and the operator product expansion
allows for model independent predictions for exclusive and inclusive $B$
decays.  These predictions will play an important role in the determination of
the $b \rightarrow c$ and $b \rightarrow u$ elements of the
Cabibbo--Kobayashi--Maskawa matrix.  These lectures provide a rudimentary
introduction to these methods. 

Chiral perturbation theory and Lattice Monte
Carlo techniques are also useful in conjunction with the heavy quark methods
developed in my lectures. 
The latter has been discussed in other lectures in this school and  reviews
on combining heavy quark and chiral symmetries can be 
found in Refs. ~\cite{wise,casalbuoni}.

Recent advances in NRQCD have improved our understanding of the
decays and production of quarkonia.  These results 
are particulary dramatic for the the production of quarkonia at high energy
accelerators where the octet matrix elements often dominate.

NRQCD and HQET remain active areas of study and development.  Even without the
discovery of dramatic new theoretical techniques the continuing experimental
developments will provide new and challenging opportunities to apply these
effective field theories.

\end{document}